\documentclass[a4paper,12pt]{article}

\usepackage[english]{babel}

\usepackage{amsmath, amssymb, amsthm}

\usepackage{graphicx}

\usepackage{float}

\usepackage{subcaption}

\usepackage{epstopdf}

\usepackage{textcomp, gensymb}

\usepackage{url}

\makeatletter 
\g@addto@macro{\UrlBreaks}{\do\/\do\-} 
\makeatother

\usepackage{natbib}

\usepackage[hidelinks]{hyperref}

\usepackage{xcolor}

\title{Estimating the maximum possible earthquake magnitude using extreme value methodology: the Groningen case}
\author{Jan Beirlant \and Andrzej Kijko \and Tom Reynkens \and John H.J.~Einmahl}

\begin{document}

\maketitle

\begin{abstract}
\noindent The area-characteristic, maximum possible earthquake magnitude $T_M$ is required by the earthquake engineering community, disaster management agencies and the insurance industry. 
The Gutenberg-Richter law predicts that earthquake magnitudes $M$ follow a truncated exponential distribution. In the geophysical literature several estimation procedures were proposed, see for instance \citeauthor{KS} (\textit{Acta Geophys.}, \citeyear{KS}) and the references therein. Estimation of $T_M$ is of course an extreme value problem to which the classical methods for endpoint estimation could be applied. We argue that recent methods on truncated tails at high levels (\citeauthor{Truncation}, \textit{Extremes}, \citeyear{Truncation}; \textit{Electron. J. Stat.},  \citeyear{trunc_real}) constitute a more appropriate setting for this estimation problem. We present upper confidence bounds to quantify uncertainty of the point estimates. We also compare methods from the extreme value and geophysical literature through simulations. Finally, the different methods are applied to the magnitude data for the earthquakes induced by gas extraction in the Groningen province of the Netherlands.
\end{abstract}

\section{Introduction}

Under the Dutch province of Groningen lies one of the largest gas fields in the world. The reservoir lies at a depth of 3~km in Rotliegend sandstone and contains an estimated 2800 billion cubic metres of gas. Since production started in 1963, around 2000 billion cubic metres of gas has been produced up to 2012 by the NAM \textit{(Nederlandse Aardolie Maatschappij)}, a partnership between Shell and ExxonMobil. As a result of taxes and its participation in NAM, the Dutch government typically receives 70\% of the profit from the Groningen Gas Field (GGF), although in some periods this can be even as high as 90\% \citep{Groningen_env}.

Despite the economic advantages of the gas extraction on the Dutch government finances, there is also a serious drawback. Since 1986, anthropogenic (man-made) seismicity is observed in the, otherwise mostly aseismic, northern part of the Netherlands, and especially in the province of Groningen. When the gas is extracted, the porous layer of sandstone, in which it is contained, compacts. Normally, this happens gradually, and the surface subsides without causing any problem. However, when this process happens e.g.\@ close to fault lines, the sandstone layers can locally compact differently which causes seismic activity \citep{vanEck, Groningen_env}. Because of this anthropogenic seismicity, houses have been damaged and the NAM has paid around 200 million euro of compensation up to 2014. Moreover, several thousands of houses need to be reinforced to avoid serious damage caused by future potential seismic activity. \citet{vanEck} also mentions other social impacts of the seismic activity including declining house prices and concerns about breaching of the dykes in the gas field area in case of a large seismic event. 

One of the obvious parameters responsible for the damage caused by seismic activity is the magnitude of the seismic event, which is directly linked to the energy released by the seismic event. So far, the largest (local) seismic event magnitude observed in the GGF is $M=3.6$ which occurred on 16 August 2012 near the village of Huizinge, municipality of Loppersum. A Modified Mercalli Intensity of VI was observed less than 4~km from the event epicentre \citep{KNMI_Huizinge}. The event caused significant damage to the infrastructure. 

A natural question arises: what is the maximum possible seismic event magnitude $T_M$ which can be generated by the GGF? Knowledge of this parameter is required by the local authorities, the engineering community, disaster management agencies, environmentalists, and the insurance industry. Its value depends on the regional tectonic setting of the area, the presence of active (capable) tectonic faults and, up to certain extent, the production regime. According to a comprehensive study of anthropogenic seismicity since 1929, the largest observed seismic event magnitude caused by oil and gas extraction is 7.3 \citep{Davies}. This event and another event of magnitude $\sim$7.0 took place near Gazli in Uzbekistan, in an area that is known to be aseismic. At the Lacq gas field in France, an event of magnitude $\sim$6.0 was recorded \citep{Lacq}.  It is uncertain that the events were indeed of anthropogenic origin, but several factors suggest that these are  examples of the strongest seismic events related to gas extraction from the gas fields. Seismicity generated by groundwater extraction has a similar character. On 11 May 2011, in Lorca, Spain, extensive groundwater extraction caused the occurrence of a shallow (2-4~km) seismic event of magnitude 5.1, leading to nine casualties and significant damage to infrastructure \citep{Lorca}. 

The purpose of this research is to assess the maximum possible seismic event magnitude $T_M$, based on the available seismic event catalogue of anthropogenic seismicity generated by the GGF. Several such estimates for the area have been made by the KNMI \textit{(Koninklijk Nederlands Meteorologisch Instituut)}: 3.3 in 1995, 3.8 in 1998 and 3.9 in 2004. In March 2016, a workshop was held in Amsterdam to provide an estimate for the maximum possible seismic event magnitude, which can be generated by the GGF (see \citet{NAM} for an overview of the results). The range of $T_M$ estimates, provided by the experts, is 3.8 to 5.0. So far, the epicentres of all occurred seismic events are within the areas of the gas extraction or not more than 500~m outside of the extraction area. This indicates that the observed seismicity can be classified as anthropogenic. However, it cannot be excluded that in the future the stresses generated by the gas extraction will be able to trigger tectonic origin stresses, resulting in significantly stronger events outside of the gas field. As a rule, such events can be significantly stronger than purely induced \citep[see e.g.\@][]{Gibowicz_Kijko}. So far, experts have found no evidence that the Groningen gas fields are capable of triggering significantly stronger seismicity than already observed. However, if such events would occur, experts believe that an event of magnitude at most 7.25 can take place \citep{NAM}.

The estimation of $T_M$ can be done in many different ways. For a review of different methods applicable for the assessment of $T_M$, see e.g.\@ \citet{KijkoGraham, Kijko, Wheeler, KS, VermeulenKijko}. A comprehensive discussion of $T_M$ assessment techniques, mainly related and applicable to fluid injection, is provided in \citet{Yeck}. Unlike \citet{Shapiro} and \citet{Hallo}, \citet{Yeck} assumed that the parameters describing the anthropogenic seismic regime (seismic activity rate, the $b$-value of Gutenberg-Richter and an upper limit of magnitude $T_M$) are subject to significant spatial and temporal variation. Especially prone to time-space fluctuation is the value of $T_M$. \citet{Yeck} suggests two different approaches for the assessment of this parameter. The first one is based on the observation \citep[see e.g.\@][]{McGarr76, McGarr14, McGarr02, Nicol} that the maximum seismic event magnitude is linearly proportional to the logarithm of the cumulative volume of fluid injected/extraction. However, \citet{Yeck} is not answering the question about the saturation of such a time dependence plot. Since fault sizes are limited, and the seismic event magnitude is linked to the fault size, the magnitudes also need to have an upper limit. Based on this simple physical consideration, the $T_M$ value must reach this certain upper limit. The second approach to assess $T_M$, which is explored in \citet{Yeck}, is based on the relationship between the size of the fault rupture and the seismic event magnitude \citep[see e.g.\@][]{Wells,Stirling}. A similar approach, extended by application of the logic tree formalism, is suggested in \citet{Bommer}. The drawback of the proposed method is the fact that anthropogenic seismicity often takes place in previously inactive areas with unknown and unmapped faults. 

Clearly, assessment of the upper limit of magnitude $T_M$ can be done using statistical tools, in particular extreme value theory (EVT). In this work, the EVT formalism is applied for assessment of the maximum possible seismic event magnitude in the GGF, by application of two different parameter estimation techniques, as developed in \citet{Truncation,trunc_real}. Our work also includes analyses of the confidence bounds of the upper limit of the magnitude distribution. For this purpose, we applied the asymptotic techniques as developed in \citet{Truncation,trunc_real}. 
Other EVT-based estimators using the moment estimator \citep{Dek} or the peaks-over-threshold maximum likelihood (POT-ML) approach have also been applied \citep[see e.g.\@][]{SoE,dHF}. However, comprehensive tests based on simulated data show that moment and POT-ML based estimators perform worse for truncated distributions than the estimators developed in \citet{Truncation,trunc_real}. For this reason, the moment and the POT-ML endpoint estimators are not discussed in this work. 
Recently, another endpoint estimator based on EVT was proposed in \citet{FANR2017}. It is however not suitable to estimate the endpoint when the distribution is truncated, as is for example the case for the Gutenberg-Richter distribution we discuss later. When applying the estimator to simulated data or the GGF data example, which we consider later in this paper, the method of \citet{FANR2017} yields very volatile estimates. Therefore, it is not included in this paper.
\newpage
The EVT-based estimators of the upper limit of the earthquake magnitudes, or equivalently the endpoint of the magnitude distribution, have received rather limited attention in the respectable seismological literature. \citet{Pisarenko_Mmax,Pisarenko_Mmax2,PisarenkoRodkin,VermeulenKijko} are notable exceptions. 

Based on empirical evidence, it is often assumed that earthquake magnitudes follow the so-called Gutenberg-Richter (GR) distribution \citep{GR}. The original GR magnitude distribution has no upper limit. After right truncation of the GR distribution, or physically speaking, after introducing the upper limit of seismic event magnitude \citep{Hamilton,Page}, the cumulative distribution function (CDF) takes the form 
\[F_M(m)=\mathbb{P}(M\leq m)=\begin{cases}
0 & \text{if } m \leq t_M\\
\frac{\exp(-\beta t_M)-\exp(-\beta m)}{\exp(-\beta T_M)-\exp(-\beta m)} & \text{if } t_M< m < T_M\\
1 &  \text{if } m \geq T_M,
\end{cases}
\]
where $t_M > 0$ is the level of completeness of the seismic event catalogue, $T_M$ is the maximum possible magnitude, i.e.\@ the upper limit (truncation point) of the magnitude distribution, and $\beta > 0$ the distribution parameter. Note that the Gutenberg-Richter distribution is not only derived empirically. There are several attempts \citep[see e.g.\@][]{Scholz,Scholz2,Rundle} to derive the GR relation based on the physical principles of earthquake occurrence or by application of the universal concept of entropy \citep[see e.g.\@][]{BerrilDavis}. Several parametric estimators of $T_M$ have been derived, which are based on the GR magnitude distribution \citep[see e.g.\@][]{Pisarenko,Pisarenko_Mmax,Raschke}. Here, we only look at one parametric estimator of $T_M$: the Kijko-Sellevoll estimator \citep{Kijko_Sellevoll,Kijko}. Moreover, we are also analysing a parametric upper confidence bound for $T_M$ based on the truncated GR distribution \citep{Pisarenko91}. This technique is applied in \citet{Potsdam} to assess the maximum possible seismic event magnitude in the GGF. Note that Bayesian estimators for the maximum earthquake magnitude have also been considered, see e.g.\@ \citet{Cornell_Bayesian,Holschneider_Bayesian, Kijko_Bayesian}.

Another parametric model for earthquake magnitudes is the tapered Pareto distribution a.k.a.\@ the modified GR distribution \citep[see e.g.\@][]{KaganJackson, KaganSchoenberg}. However, unlike the truncated GR distribution, this model does not provide an upper bound for the magnitudes, which makes it unrealistic from a physical point of view.

\citet{Potsdam} like e.g.\@ \citet{PisarenkoRodkin} provides estimates for the maximum expected seismic event magnitude to occur, for different time intervals (time horizons). It is important to note that in our work, we do not try to estimate that quantity, but we only look at estimates for the time-independent maximum possible seismic event magnitude.

In the next section, we discuss the different endpoint estimators that can be applied to estimate the maximum possible seismic magnitude $T_M$. In Section 3, we apply these methods to estimate $T_M$ for the GGF. Moreover, we also discuss upper confidence bounds for $T_M$. Afterwards, we compare the performance of the EVT-based estimators with some discussed in \citet{KS} using simulations, assuming that the seismic event magnitude is distributed according to the truncated GR distribution.

\section{Overview of applied estimators}

We now discuss several different types of endpoint estimators: the EVT-based estimators are presented in Section~\ref{sec:endpointEVT}, the non-parametric estimators as discussed in \citet{KS} are described in Section~\ref{sec:endpointNP} and the parametric Kijko-Sellevoll estimator is presented in Section~\ref{sec:endpointKS}. We provide only very few details for the estimators already in use for assessment of the upper limit of the seismic event magnitude. More details can be found in \citet{Kijko, KS}. In all cases where order statistics are used, the ordered sample of magnitudes is denoted as $M_{1,n} \leq \ldots \leq  M_{n,n}$.

\subsection{EVT-based estimators}\label{sec:endpointEVT}

We consider two EVT-based estimators of the endpoint: the truncated generalised Pareto distribution (GPD) estimator using the framework from \citet{trunc_real} and the truncated Pareto estimator of \citet{Truncation}.

The methodology for modelling the upper tail of the distribution of a random variable $Y$ relies on the fact that the maximum of independent measurements $Y_i, \; i=1,\ldots,n,$  can be approximated by the generalised extreme value distribution: as $n\to \infty$,
\begin{equation}
\mathbb{P}\left(\frac{\displaystyle\max_{i=1,\ldots,n}Y_i -b_n}{a_n} \leq y \right) \to
G_{\xi} (y) = \exp \left( - (1 + \xi y)^{-1/\xi} \right),   \;\; 1+\xi y>0,
\label{eq:maxd}
\end{equation}
where $b_n \in \mathbb{R}$, $a_n >0$ and $\xi \in \mathbb{R}$ are the location, scale and shape parameters, respectively. For $\xi =0$, $G_0(y)$ has to be read as $\exp\left(- \exp (-y)\right)$.
In fact, \eqref{eq:maxd} represents the only possible non-degenerate limits for maxima of independent and identically distributed sequences $Y_i$. Let
$F_Y (y) = \mathbb{P}(Y \leq y)$ denote the CDF, $\bar{F}_Y(y) = 1-F_Y(y)$ the right tail function (RTF),  and  $Q_Y (p) = \inf \{y  \,|\, F_Y(y) \geq p \}$ ($0<p<1$) the quantile function of a random variable $Y$. 
\par Condition \eqref{eq:maxd} is equivalent to the convergence of the distribution of excesses (or peaks) over high thresholds $t$ to the generalised Pareto distribution (GPD):
as $t$ tends to the endpoint of the distribution of $Y$, then,
\begin{equation}
	\mathbb{P} \left(\frac{Y -t}{\sigma_t}>y \,\middle|\, Y>t\right) =
	\frac{\bar{F}_Y (t+y \sigma_t )}{\bar{F}_Y (t)}
	\to H_{\xi}(y) = -\ln G_{\xi} (y) = \left( 1+\xi y\right)^{-1/\xi},
	\label{eq:pot}
	\end{equation}
where $\sigma_t >0$. The shape parameter $\xi$ is often called the extreme value index (EVI). 
The specific case $\xi >0$ consists of the Pareto-type distributions defined through
\begin{equation}
\frac{Q_Y(1-\frac{1}{vy})}{Q_Y(1-\frac{1}{y})} \to_{y\to \infty} v^{\xi} \quad \mbox{ and } \quad
\mathbb{P} \left(\frac Yt>y \,\middle|\, Y>t\right) = \frac{\bar{F}_Y (ty )}{\bar{F}_Y (t)}
\to_{t\to \infty} y^{-1/\xi}.
\label{eq:Pa}
\end{equation}
The max-domain of attraction (MDA) in case $\xi=0$  is called the Gumbel domain to which exponentially decreasing tails belong. Finally, the domain corresponding to negative values of the EVI have finite right endpoints. 

Right truncation models for $X$ based on a parent variable $Y$ satisfying the above extreme value assumptions, are obtained from
\begin{equation}
X =_d (Y\,|\,Y<T),
\label{eq:T}
\end{equation}
for some $T>0$.
The odds of the truncated probability mass under the untruncated distribution $Y$ are denoted by $D_T = \bar{F}_Y (T)/F_Y(T)$.

Truncation with the threshold $t=t_n \to \infty$ is  defined through the assumption
\begin{equation}
\frac{T-t}{\sigma_t} \to \kappa >0,
\label{eq:C}
\end{equation}
which then entails that for $x\in (0,\kappa)$
\begin{equation}
\mathbb{P}\left(\frac{X-t}{\sigma_t} >x \,\middle|\, X>t\right) \to \frac{(1+ \xi  x)^{-1/\xi}  -  (1+ \xi \kappa )^{-1/\xi}}
{1-(1+ \xi \kappa)^{-1/\xi}}.
\label{eq:FT}
\end{equation}
This corresponds to situations where  the deviation from the GPD behaviour due to truncation at a high value $T$ will be visible in the data from $t$ on, and the approximation of the peaks-over-threshold (POT) distribution using the limit distribution in \eqref{eq:FT} appears more appropriate than with a simple GPD. 

In the specific case of Pareto-type distributions (i.e.\@ $\xi >0$) condition \eqref{eq:FT} can be simplified to
\begin{equation}
\mathbb{P}\left(\frac{X}{t} >x \,\middle|\, X>t\right) \to \frac{x^{-1/\xi} - \rho^{-1/\xi}}{1- \rho^{-1/\xi}}, \;\; 1< x < \rho,
\label{eq:FT+}
\end{equation}
assuming that
$T/t \to \rho > 1$.

In practice, one has to choose a certain threshold $t$. Often, one takes it equal to the $(k+1)$-th largest observation $X_{n-k,n}$ and then computes the estimator for many values of $k$.

\subsubsection{Truncated GPD estimator}

We can estimate the endpoint of the magnitude distribution using the techniques developed in \citet{trunc_real}. Its estimator for the truncation point $T_M$ is based on condition \eqref{eq:FT} for the variable $M$ where $\xi$ is the EVI of $Y$, the parent variable of $M$, see Table~\ref{tab:notation_magen}. The corresponding estimator for the endpoint is then given by
\begin{equation}
\hat{T}^M_{k} = M_{n-k,n} + \frac{1}{\hat\tau_k}\left[\left( \frac{1-\frac1k}{(1+ \hat\tau _k (M_{n,n}-M_{n-k,n}))^{-1/\hat\xi_k}-\frac1k} \right)^{\hat\xi _k} -1 \right],
\label{eq:hatTmle}
\end{equation} 
with $\hat{\xi}_k$ and $\hat{\tau}_k$ the estimates for $\xi$ and $\tau = \xi/\sigma_t$ obtained by application of the maximum likelihood principle. See \citet{trunc_real} for more details on estimation and testing.
We will call this estimator the \textit{Truncated GPD}.

\par Using Theorem~2 in \citet{trunc_real} with $p=0$, we obtain an approximate \mbox{$100(1-\alpha)\%$} upper confidence bound for $T_M$:
\begin{equation}\label{eq:CI_trMLE}
\hat{T}^M_{k} -(\ln\alpha+1)\frac{\frac{k+1}{(n+1)\hat{D}_{T,k}}}{k+1}\left(1+\frac{k+1}{(n+1)\hat{D}_{T,k}}\right)^{\hat{\xi}_k}\frac{\hat{\xi}_k}{\hat{\tau}_k}\, .
\end{equation}
One has to note that in \eqref{eq:CI_trMLE}, second order terms have been omitted, and $\hat{D}_{T,k}$ denote the estimates for the truncation odds $D_T$, see \citet{trunc_real}.

\begin{table}[ht]
	\centering
	\resizebox{\textwidth}{!}{  
		\begin{tabular}{c|cc|cc}
			\hline
			Variable & EVI & Endpoint & Parent variable & EVI of parent variable \\
			\hline
			Magnitude $M$ & $\xi_M$ & $T_M$ & $Y$ with $M =_d (Y \, | \, Y<T_M)$ & $\xi$ \\
			Energy $E$ & $\xi_E$ & $T_E$ & $Y_E$ with $E=_d (Y_E \, | \, Y_E<T_E)$ & $\xi_{Y_E}$ \\
			\hline
	\end{tabular}}%
	\caption{Magnitude and energy: overview of notation.}\label{tab:notation_magen}
\end{table}

\subsubsection{Truncated Pareto estimator}\label{sec:trHill}

The endpoint estimator of \citet{Truncation} is based on condition \eqref{eq:FT+} and is hence only suitable for truncated Pareto-type tails. Since the (truncated) GR magnitude distribution is a truncated exponential distribution, we expect that this estimator cannot be applied to the magnitudes directly. Instead, we use following empirical relationship between the (local) earthquake magnitude $M$ and the energy released by earthquakes \citep{MGS}  
\begin{equation}
E = 2\times 10^{1.5(M-1)} = \exp(\ln2+(M-1)1.5\ln10),
\label{eq:energy_mag}
\end{equation}
or reversely
\begin{equation}
M = \frac{\log_{10} \left( \frac{E}2\right)}{1.5}+1=\frac{\ln \left( \frac{E}2\right)}{1.5\ln10}+1,
\label{eq:mag_energy}
\end{equation}
where the energy is expressed in megajoules (MJ). We thus expect the energy to follow a truncated Pareto-type distribution. Therefore, we apply the estimator of \citet{Truncation} to the energy and transform the endpoint back to the magnitudes using \eqref{eq:mag_energy}. By denoting the parent variable of $E$ by $Y_E$, we have \mbox{$ E=_d (Y_E \,|\, Y_E<T_E)$}, where $T_E$ is the endpoint for $E$, see Table~\ref{tab:notation_magen}. 

\par Using the approach of \citet{Truncation} applied to the variable $E$, the endpoint for the energy is then estimated as 
\begin{equation}\label{eq:endpoint_trHill_energy}
\hat{T}^{E,+}_{k} = 2\times 10^{1.5(M_{n-k,n}-1)} \left(\frac{\left(\frac{2\times 10^{1.5(M_{n-k,n}-1)}}{2\times 10^{1.5(M_{n,n}-1)}}\right)^{1/\hat{\xi}^{Y_E,+}_k}-\frac1{k+1}}{1-\frac1{k+1}}\right)^{-\hat{\xi}^{Y_E,+}_k}.
\end{equation}
Here, $\hat{\xi}^{Y_E,+}_k$ are the estimates for $\xi_{Y_E}$, the extreme value index of $Y_E$. See \citet{Truncation} for more details on estimation and testing. Note that $\rho$ (see \eqref{eq:FT+}) is estimated by $E_{n,n}/E_{n-k,n}$. 

Transforming the estimated endpoints for the energy gives the following endpoint estimates for the magnitudes:
\begin{equation}\label{eq:endpoint_trHill}
\hat{T}^{M,+}_{k} = \frac{\log_{10} \left( \frac{\hat{T}^{E,+}_{k}}2\right)}{1.5} +1.
\end{equation}
We denote this estimator by \textit{Truncated Pareto}.
\par Using the asymptotic results in \citet{Truncation}, an approximate $100(1-\alpha)\%$ upper confidence bound for $T_E$ can be constructed.
Application of Theorem~2 in \citet{Truncation}, after omitting second-order terms again, gives the following approximate \mbox{$100(1-\alpha)\%$} upper confidence bound for $T_E$:
\[ \exp\left(\ln \hat{T}^{E,+}_{k}-\frac{\frac{k+1}{(n+1)\hat{D}^{E,+}_{T,k}}\hat{\xi}^{Y_E,+}_k}{k+1}(\ln\alpha+1)\right).\]
Here, $\hat{D}^{E,+}_{T,k}$ are the truncated Pareto estimates for the truncation odds $D_T^E= \bar{F}_{Y_E} (T_E)/F_{Y_E}(T_E)$, see \citet{Truncation}. 
This upper bound can then be transformed back to the magnitude level as before to get an approximate $100(1-\alpha)\%$ upper confidence bound for $T_M$:
\begin{equation}\label{eq:CI_trHill}
\frac{\ln\left(\frac{\hat{T}^{E,+}_{k}}2\right)}{1.5\ln10} + 1 -\frac{\frac{\frac{k+1}{(n+1)\hat{D}^{E,+}_{T,k}}\hat{\xi}^{Y_E,+}_k}{k+1}(\ln\alpha+1)}{1.5\ln10}=\hat{T}^{M,+}_{k} -\frac{\frac{\frac{k+1}{(n+1)\hat{D}^{E,+}_{T,k}}\hat{\xi}^{Y_E,+}_k}{k+1}(\ln\alpha+1)}{1.5\ln10}.
\end{equation}

\subsection{Non-parametric estimators}\label{sec:endpointNP}

The next estimators are all based on the fact that 
\begin{equation}\label{eq:endpoint_E}
\mathbb{E}(M_{n,n}) = \int_{t_M}^{T_M} m \,d F_M^n(m)= T_M - \int_{t_M}^{T_M} F_M^n(m) \, dm,
\end{equation}
see \citet{KS}.
Hence, $T_M$ can be estimated by
\[\hat{T}^M= M_{n,n} + \Delta\]
with $\Delta$ an estimator for $\int_{t_M}^{T_M} F_M^n(m) \, dm$.

\subsubsection{Non-parametric with Gaussian kernel}

The CDF in \eqref{eq:endpoint_E} can be estimated using a Gaussian kernel.
The estimator for the endpoint is then obtained as the iterative solution of the equation
\begin{equation*}\label{eq:endpoint_npG1}
T_M = M_{n,n} +\Delta
\end{equation*}
with
\begin{equation}\label{eq:endpoint_npG2}
\Delta = \int_{t_M}^{T_M} \left(\frac{\sum_{i=1}^n \Phi\left(\frac{m-M_i}h\right)-\Phi\left(\frac{t_M-M_i}h\right)}{\sum_{i=1}^n \Phi\left(\frac{T_M-M_i}h\right)-\Phi\left(\frac{t_M-M_i}h\right)}\right)^n \,dm
\end{equation}
and $\Phi$ the standard normal CDF.
The bandwidth $h$ is chosen using unbiased cross-validation. 
We denote this estimator by \textit{N-P-G}.  For more details we refer to \citet{Kijko2} and Equations 28 and 29 in \citet{KS}.

\subsubsection{Non-parametric based on order statistics}
\citet{Cooke79} proposes to approximate the CDF in \eqref{eq:endpoint_E} with the empirical CDF.
The corresponding endpoint estimator, see Equation 33 in \citet{KS}, is given by
\begin{equation}\label{eq:endpoint_npOS}
\hat{T}_n^{M,N-P-OS} = M_{n,n} + \left[ M_{n,n}-(1-\exp(-1)) \sum_{i=0}^{n-1} \exp(-i) M_{n-i,n}\right].
\end{equation}
We denote this estimator by \textit{N-P-OS}.
\par \citet{Cooke79} also constructed an approximate $100(1-\alpha)\%$ upper confidence bound for $T_M$:
\begin{equation}\label{eq:CI_npOS}
M_{n,n} + \frac{M_{n,n}-M_{n-1,n}}{(1-\alpha)^{-\nu}-1},
\end{equation}
where the parameter $\nu$ is determined by
\begin{equation*}
\lim_{y\uparrow 0} \frac{1-F_M(T_M+cy)}{1-F_M(T_M+y)}=c^{1/\nu}
\label{eq:nu_Cooke}
\end{equation*}
for every constant $c>0$. 

Note that $\nu=1$ for upper truncated distributions which can be proved by application of the mean value theorem. Since it is often assumed that magnitude data come from an upper truncated distribution, e.g.\@ the truncated Gutenberg-Richter distribution, we use $\nu=1$ in the remainder.

\subsubsection{Few largest observations}
Later, \citet{Cooke80} proposed a simple estimator that only uses the maximum and the \mbox{$(k+1)$}-th largest magnitude. This estimator, see Equation 38 in \citet{KS}, is equal to
\begin{equation}\label{eq:endpoint_FL}
\hat{T}^{M,FL}_{k} = M_{n,n} + \left[\frac1k(M_{n,n}-M_{n-k+1,n})\right].
\vspace{-0.05cm}
\end{equation}
We denote this estimator by \textit{FL}.

\subsubsection{Extended FL}
The previous estimator only uses two observations. It can be extended as
\begin{equation}\label{eq:endpoint_EFL}
\hat{T}^{M,EFL}_{k} = M_{n,n} + \left[\frac1{k} \left(M_{n,n} - \frac1{k-1} \sum_{i=2}^k M_{n-i+1,n}\right)\right],
\end{equation}
see Equation 40 in \citet{KS}. We denote this estimator by \textit{EFL}.

\subsubsection{Robson -- Whitlock}
\citet{RW} proposes the following simple estimator:
\begin{equation}\label{eq:endpoint_RW}
\hat{T}_2^{M,R-W} = M_{n,n} +\left[ M_{n,n} - M_{n-1,n}\right],
\end{equation}
see Equation 42 in \citet{KS}. We denote this estimator by \textit{R-W}.
\par Another approximate $100(1-\alpha)\%$ upper confidence bound for $T_M$ was derived in \citet{RW}:
\begin{equation}\label{eq:CI_RW}
M_{n,n} + \frac{1-\alpha}{\alpha}\left(M_{n,n}-M_{n-1,n}\right).
\end{equation}
Note that this corresponds to the upper confidence bound \eqref{eq:CI_npOS} of \citet{Cooke79} (with $\nu=1$).

\subsubsection{Robson -- Whitlock -- Cooke}
The previous estimator can be improved, in terms of MSE, as shown in \citet{Cooke79}.
The improved estimator is obtained as
\begin{equation}\label{eq:endpoint_RWC}
\hat{T}_2^{M,R-W-C} = M_{n,n} + \left[\frac1{2\nu}(M_{n,n} - M_{n-1,n})\right],
\end{equation}
see Equation 46 in \citet{KS}. As before, we take $\nu$ equal to 1. We denote this estimator by \textit{R-W-C}.
Note that this estimator corresponds to the FL estimator for $k=2$.

\subsection{Parametric estimator: Kijko -- Sellevol}\label{sec:endpointKS}
\citet{Kijko_Sellevoll} introduced the equation (see Equation 13 in \citet{KS}) 
\begin{equation}\label{eq:endpoint_KS}
T_M= M_{n,n} +\left[\frac{E_1(n_2)-E_1(n_1)}{\beta \exp(-n_2)}+t_M\exp(-n)\right]
\end{equation}
with
\[n_1=\frac{n}{1-\exp(-\beta(T_M-t_M))},\quad n_2=n_1\exp(-\beta(T_M-t_M)),\]
and $E_1(z)=\int_z^{\infty} \exp(-s)/s\, ds $ the exponential integral function. 
Since these expressions depend on $T_M$, we obtain $T_M$ using an iterative procedure.
The parameter $\beta$ is estimated based on the truncated Gutenberg-Richter law using maximum likelihood, see \citet{Page} and Chapter~12 in \citet{Gibowicz_Kijko}. It is estimated iteratively using the equation
\[\frac{1}{\beta} = \overline{M}_n -t_M + \frac{(T_M-t_M)\exp(-\beta(T_M-t_M))}{1-\exp(-\beta(T_M-t_M))},\]
where $\overline{M}_n=1/n\sum_{i=1}^n M_i$ is the sample mean of $M_1,\ldots,M_n$.
Using a Taylor expansion, this becomes
\begin{equation}\label{eq:beta_Taylor}
\hat{\beta} = \hat{\beta}_0\left(1-\hat{\beta}_0\frac{(T_M-t_M)\exp(-\hat{\beta}_0(T_M-t_M))}{1-\exp(-\hat{\beta}_0(T_M-t_M))}\right)
\end{equation}
where $\hat{\beta}_0 = \frac{1}{\overline{M}_n- t_M}$ is the Aki-Utsu \citep{Aki, Utsu} estimator for $\beta$.
This approach does not use iterations and is thus preferred for computational reasons.
In each iteration step (for $T_M$), we first update the estimate of $\beta$ using \eqref{eq:beta_Taylor}, and then improve the estimate of $T_M$.
We denote this estimator of the maximum magnitude by \textit{K-S}. Note that of all discussed estimators, this is the only one that uses the truncated Gutenberg-Richter law directly.
\par Based on the truncated Gutenberg-Richter law, a parametric  $100(1-\alpha)\%$ upper confidence bound for $T_M$ can be constructed (see Equation~19 in \citet{Pisarenko91}):
\begin{equation}\label{eq:CI_GR}
t_M-\frac1{\beta}\ln\left(\frac{\exp(-\beta(M_{n,n}-t_M))-1}{\alpha^{1/n}}+1\right),
\end{equation}
where we estimate $\beta$ using the K-S method. \citet{Holschneider_Bayesian,ZH_CI} noted that the upper confidence bound as defined in \citet{Pisarenko91} is infinite if the maximum observed seismic event magnitude is larger than $t_M-\frac1{\beta}\ln(1-\alpha^{1/n})$.  For the GGF magnitude data, this happens when $\alpha\leq 0.061$. Therefore, we consider $\alpha=0.1$ in the data example and the simulations. A comprehensive discussion on this subject, including a condition on the existence of Pisarenko's original $T_M$ estimator, can be found in \citet{VermeulenKijko}.

\section{Estimation of the maximum possible seismic event magnitude generated by the GGF}\label{sec:Groningen}

\begin{figure}[!h]
		\vspace{-0.7cm}
	\centering
	\begin{subfigure}{0.45\textwidth}
		\centering
		\includegraphics[width=\textwidth,trim={5cm 0 5cm 0},clip]{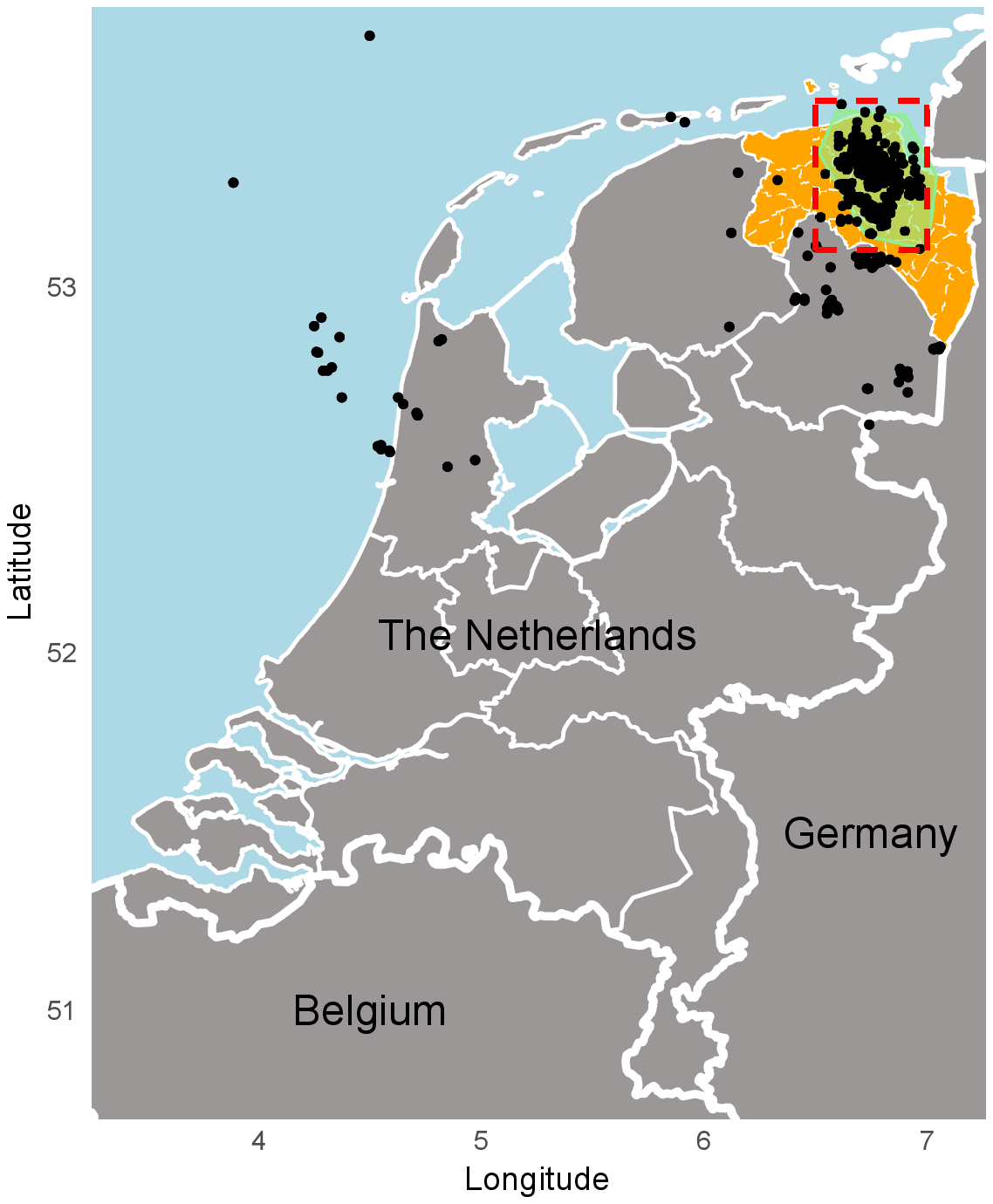}%
		\caption{}\label{fig:NL_ind}%
	\end{subfigure}%
\hfill
	\begin{subfigure}{0.45\textwidth}
		\centering
		\includegraphics[width=\textwidth,trim={5cm 0 4.75cm 0},clip]{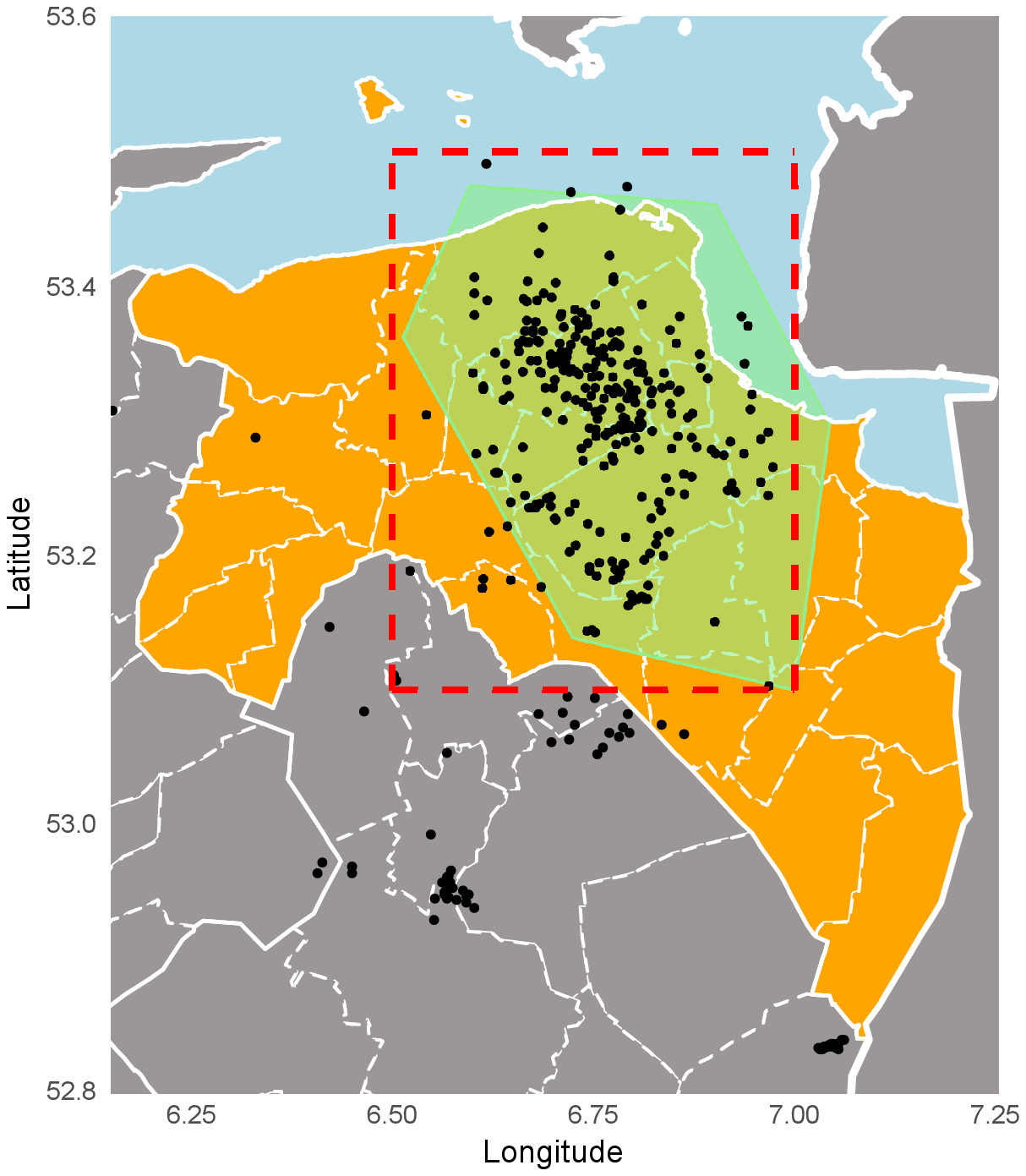}%
		\caption{}\label{fig:Groningen_ind}%
	\end{subfigure}%
	\newline
	\begin{subfigure}{0.5\textwidth}
		\centering
		\includegraphics[height=\linewidth, angle=270]{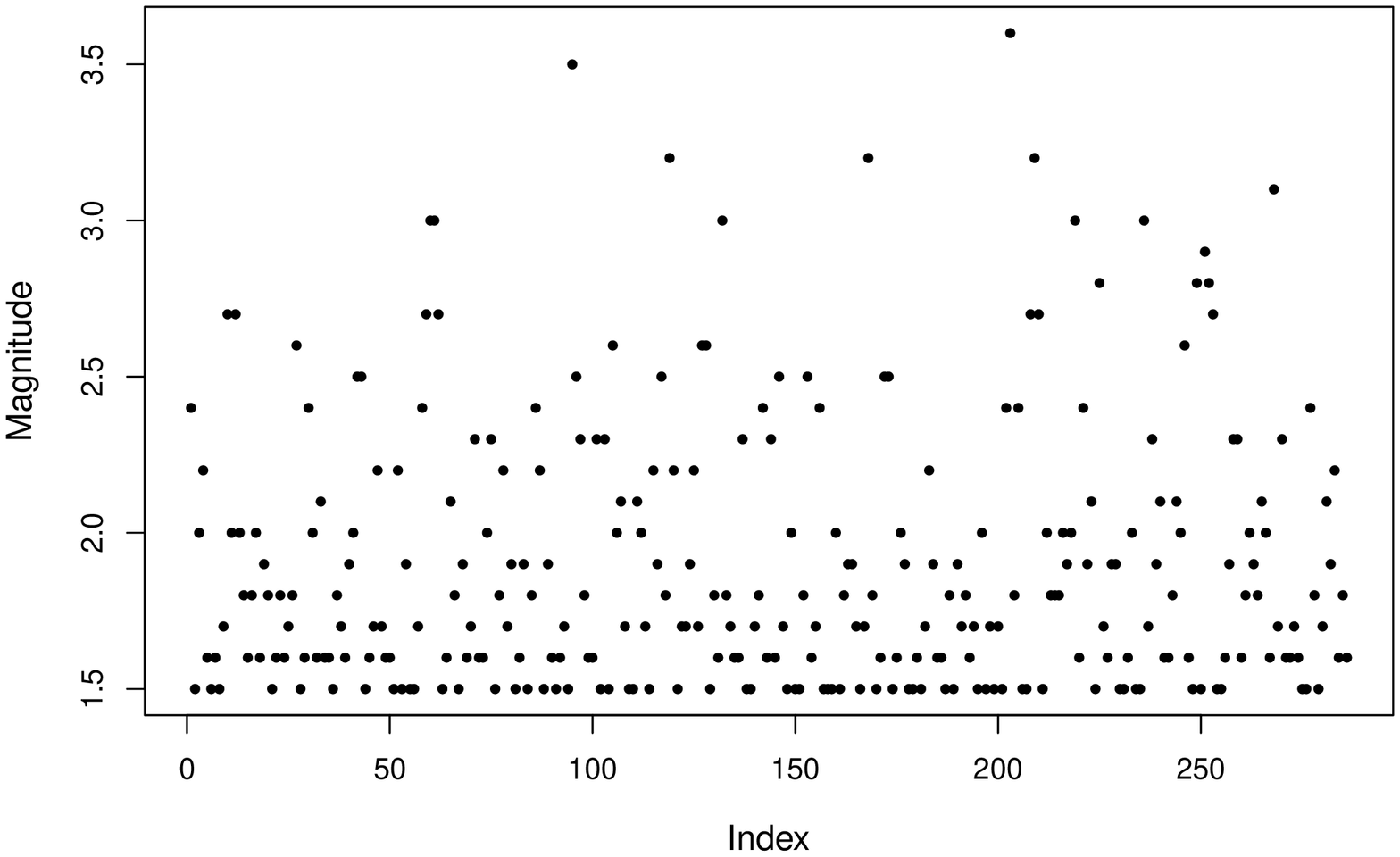}
		\caption{}\label{fig:Groningen_data}%
	\end{subfigure}
	\caption{Locations of anthropogenic seismicity in (a) the Netherlands and (b) Groningen between December 1986 and 31 December 2016 with magnitudes at least 1.5, and (c) magnitude plot of anthropogenic seismicity in the considered area with magnitudes at least 1.5.}%

\end{figure}

In this section, we attempt to estimate the maximum possible seismic event magnitude which can be generated by gas extraction in the GGF. The database of the seismicity of anthropogenic origin in the area is downloaded from the website of the KNMI: \url{https://www.knmi.nl/kennis-en-datacentrum/dataset/aardbevingscatalogus}. The database contains (local) magnitudes $M$ of seismic events of anthropogenic origin in the Netherlands.
We only consider events from the database that are located within the rectangle determined by (53.1\degree N, 6.5\degree E), (53.1\degree N, 7\degree E), (53.5\degree N, 7\degree E) and (53.5\degree N, 6.5\degree E),  see Figure~\ref{fig:NL_ind}. The selected area is almost the same as the area that was considered in \citet{Potsdam}. The extracted database contains 286 seismic events with magnitudes at least 1.5, which have been recorded between December 1986 and 31 December 2016. The events, together with the boundaries of the selected area and approximate contours of the whole GGF (green), are shown in Figure~\ref{fig:Groningen_ind}. A plot of the magnitudes of the selected events is shown in Figure~\ref{fig:Groningen_data}. The dataset was tested for serial correlation, and no significant correlation could be detected. Moreover, comparing the analysis using all earthquakes (as is done in this section) with the analysis using only more recent earthquakes did not indicate non-stationarity in the data which is also confirmed by Figure~\ref{fig:Groningen_data}.

\begin{figure}[!h]
	\centering
	\begin{subfigure}{0.495\textwidth}
		\centering
		\includegraphics[height=\textwidth, angle=270]{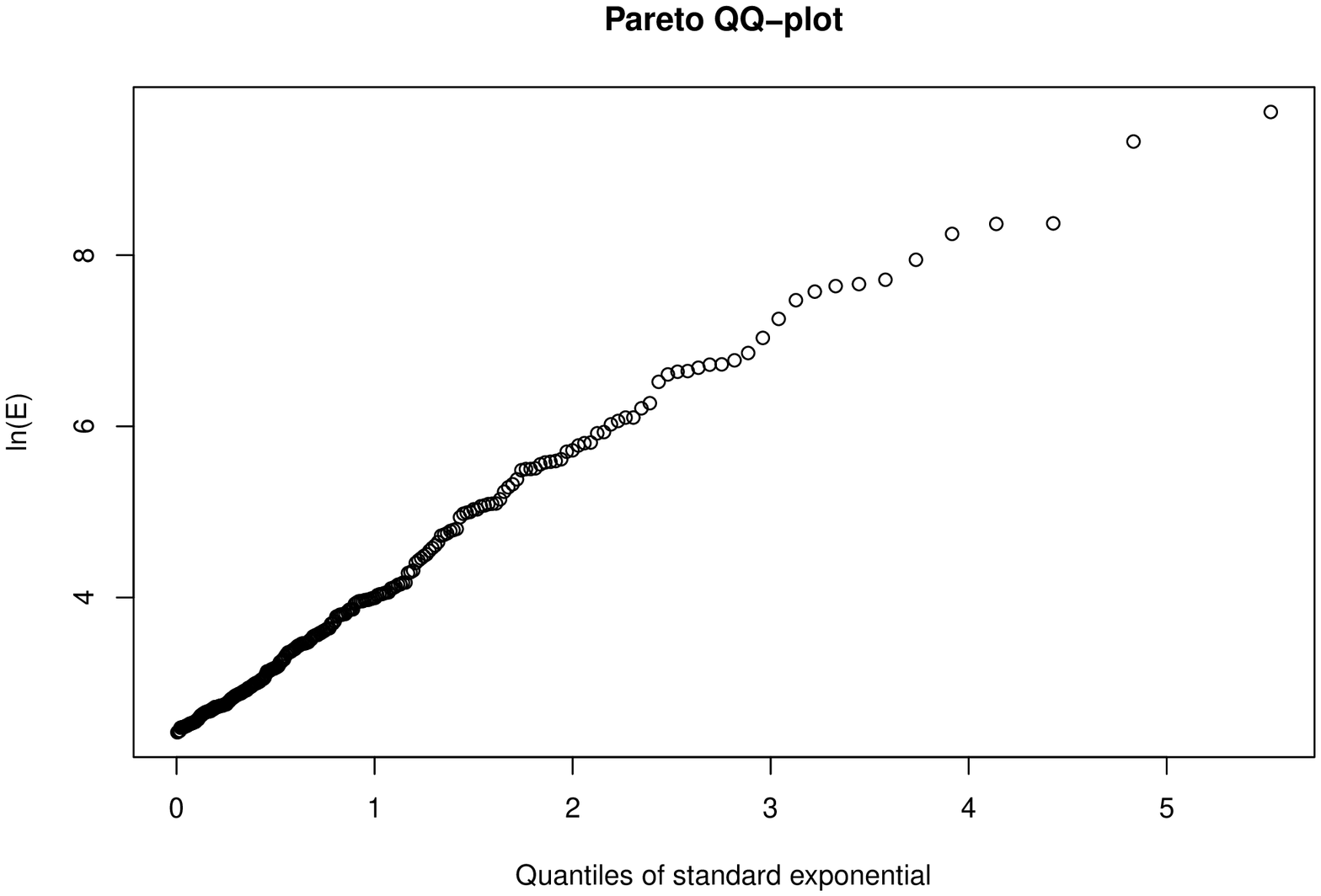}%
		\caption{}\label{fig:Groningen_ParetoQQ}%
	\end{subfigure}%
	\hfill
	\begin{subfigure}{0.495\textwidth}
		\centering
		\includegraphics[height=\textwidth, angle=270]{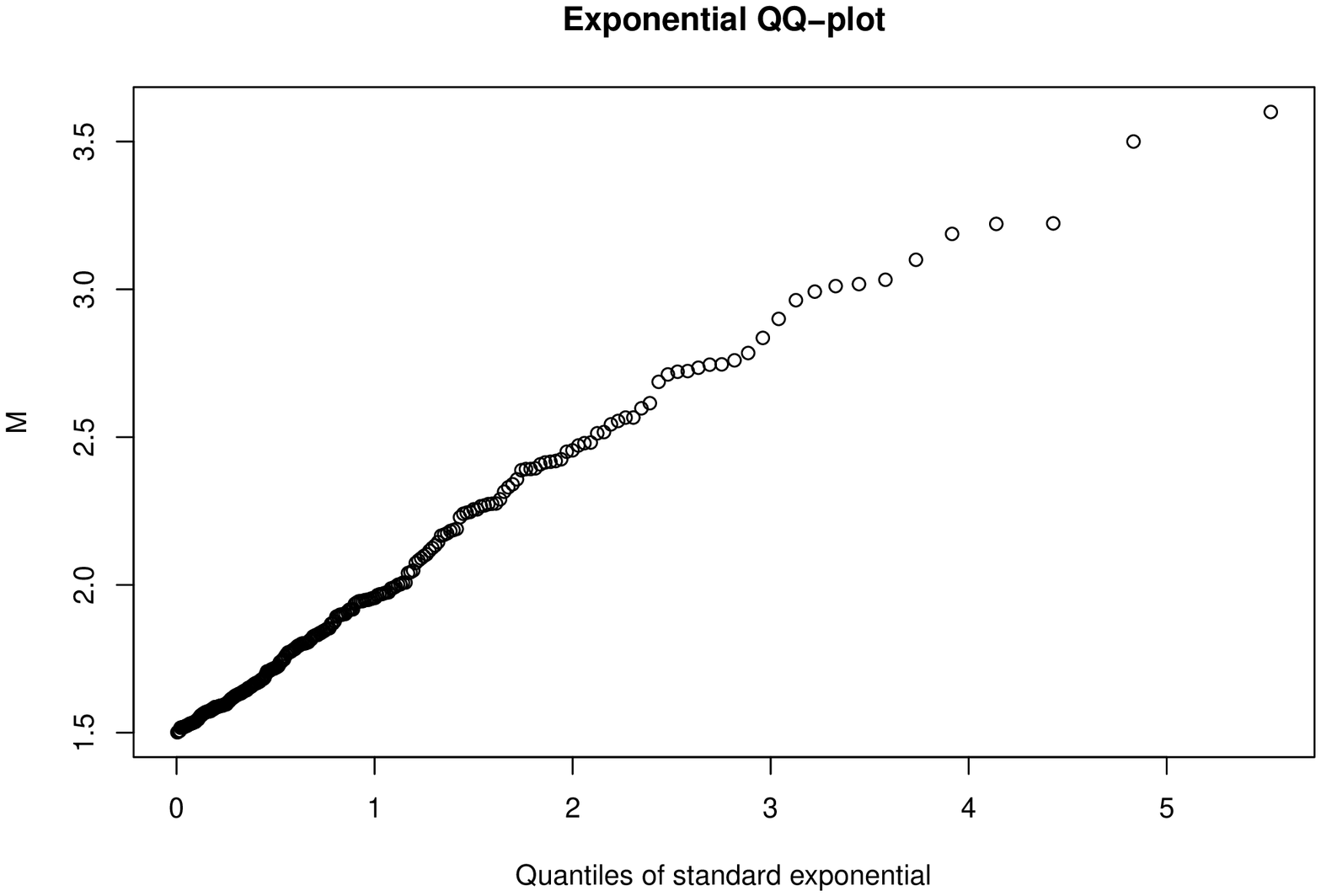}%
		\caption{}\label{fig:Groningen_expQQ}
	\end{subfigure}
	\begin{subfigure}{0.495\textwidth}
		\centering
		\includegraphics[height=\textwidth, angle=270]{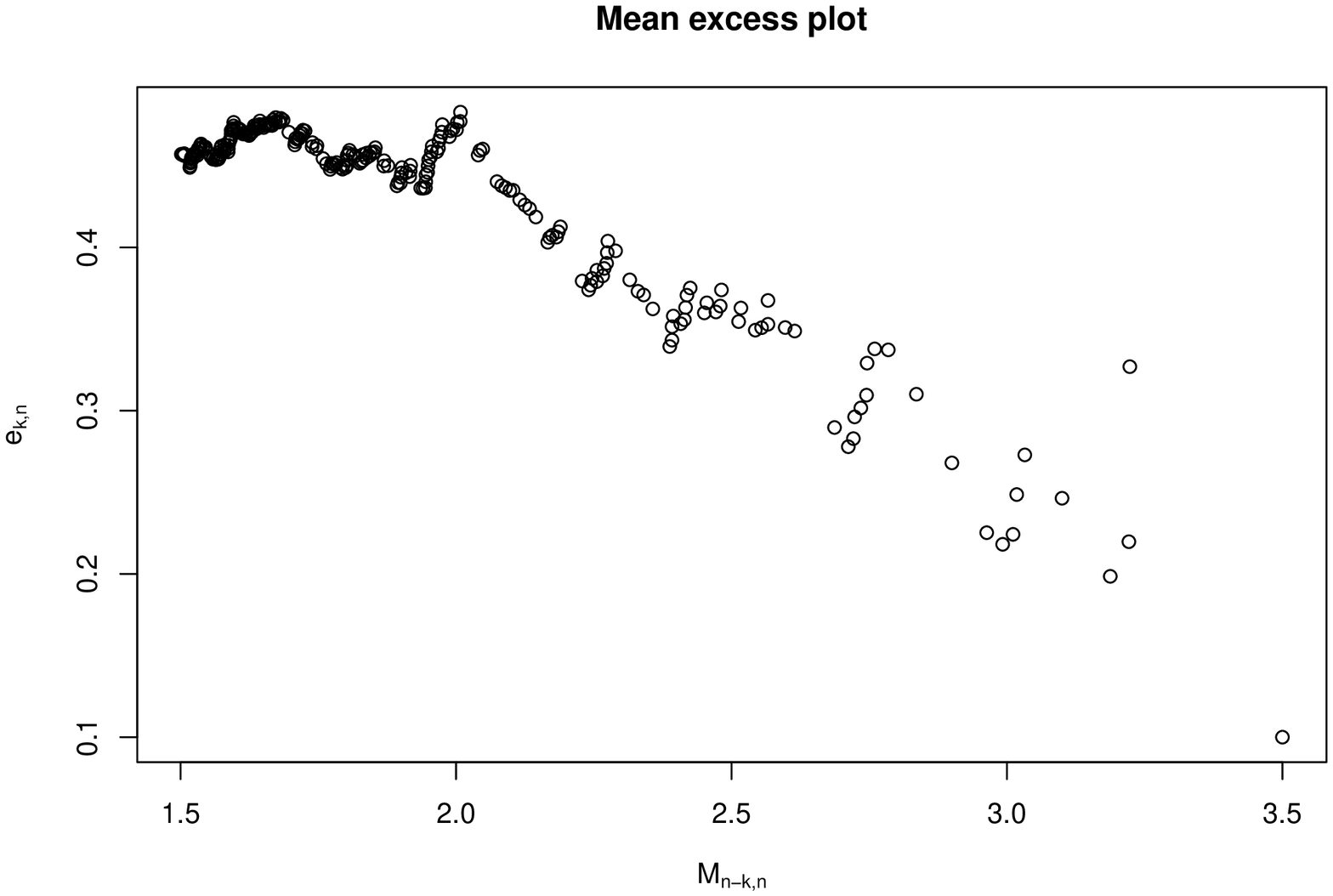}%
		\caption{}\label{fig:Groningen_MeanExcess}%
	\end{subfigure}
	\hfill
	\begin{subfigure}{0.495\textwidth}
		\centering
		\includegraphics[height=\textwidth, angle=270]{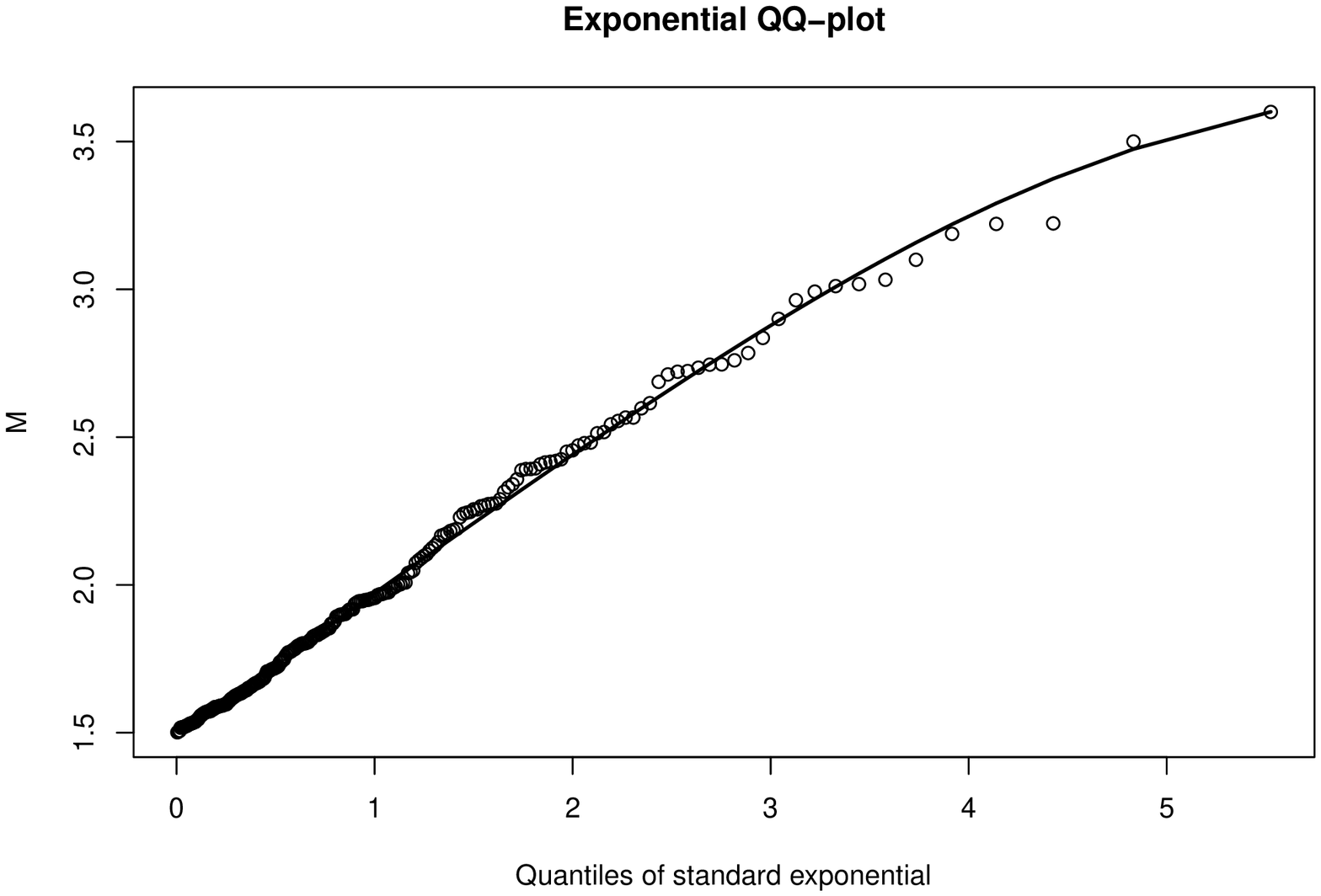}%
		\caption{}\label{fig:Groningen_ExpQQ_fit}%
	\end{subfigure}
	\caption{Groningen gas field anthropogenic seismicity: (a) Pareto QQ-plot of energy data, (b) exponential QQ-plot of magnitude data, (c) mean excess plot of magnitude data
		and (d) exponential QQ-plot of magnitude data with fit based on the $k=125$ largest magnitudes.}%
\end{figure}

\begin{figure}[!h]
	\centering
	\begin{subfigure}{0.495\textwidth}
		\centering
		\includegraphics[height=\textwidth, angle=270]{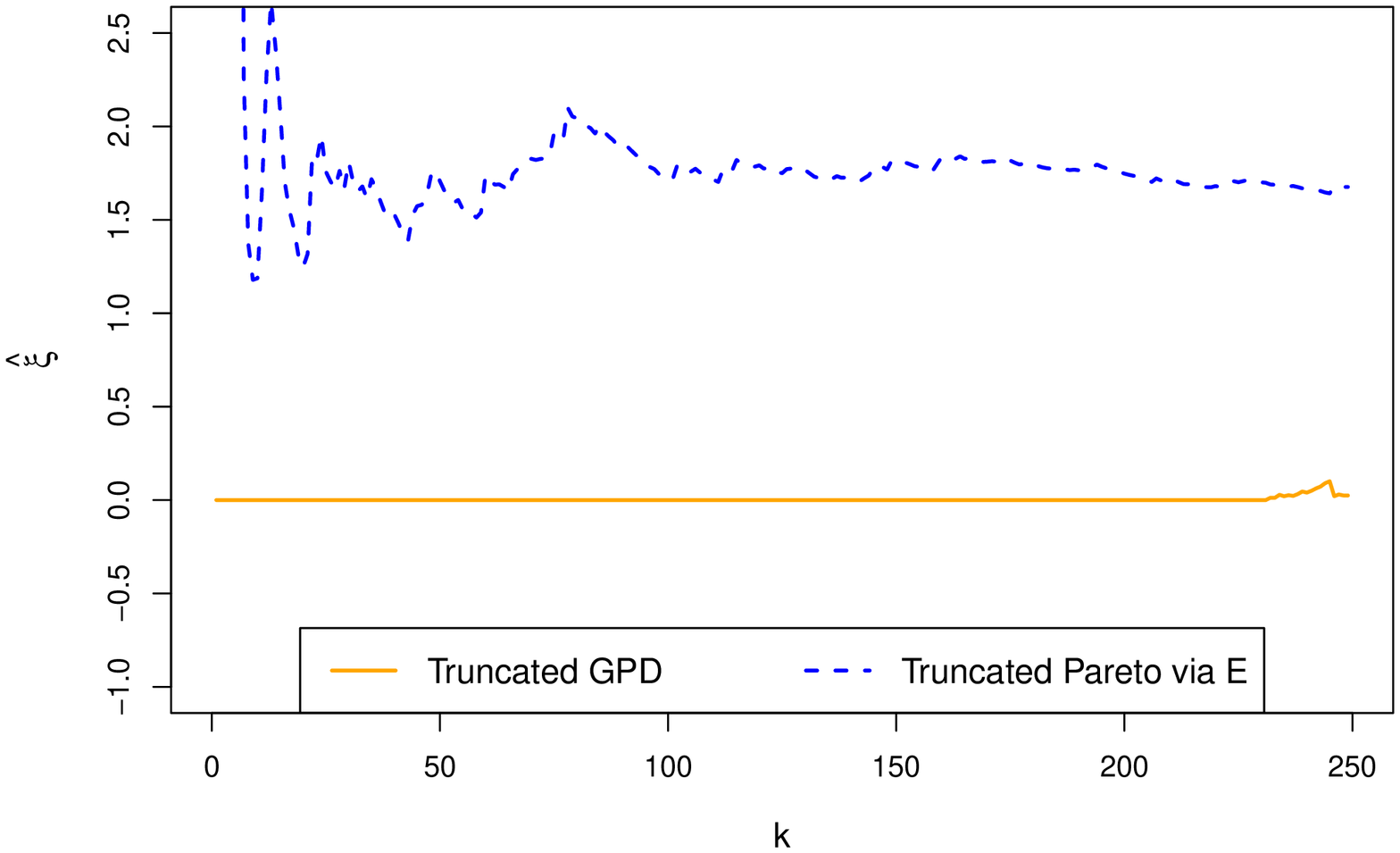}%
		\caption{}%
		\label{fig:Groningen_xi}%
	\end{subfigure}
	\hfill
	\begin{subfigure}{0.495\textwidth}
		\centering
		\includegraphics[height=\textwidth, angle=270]{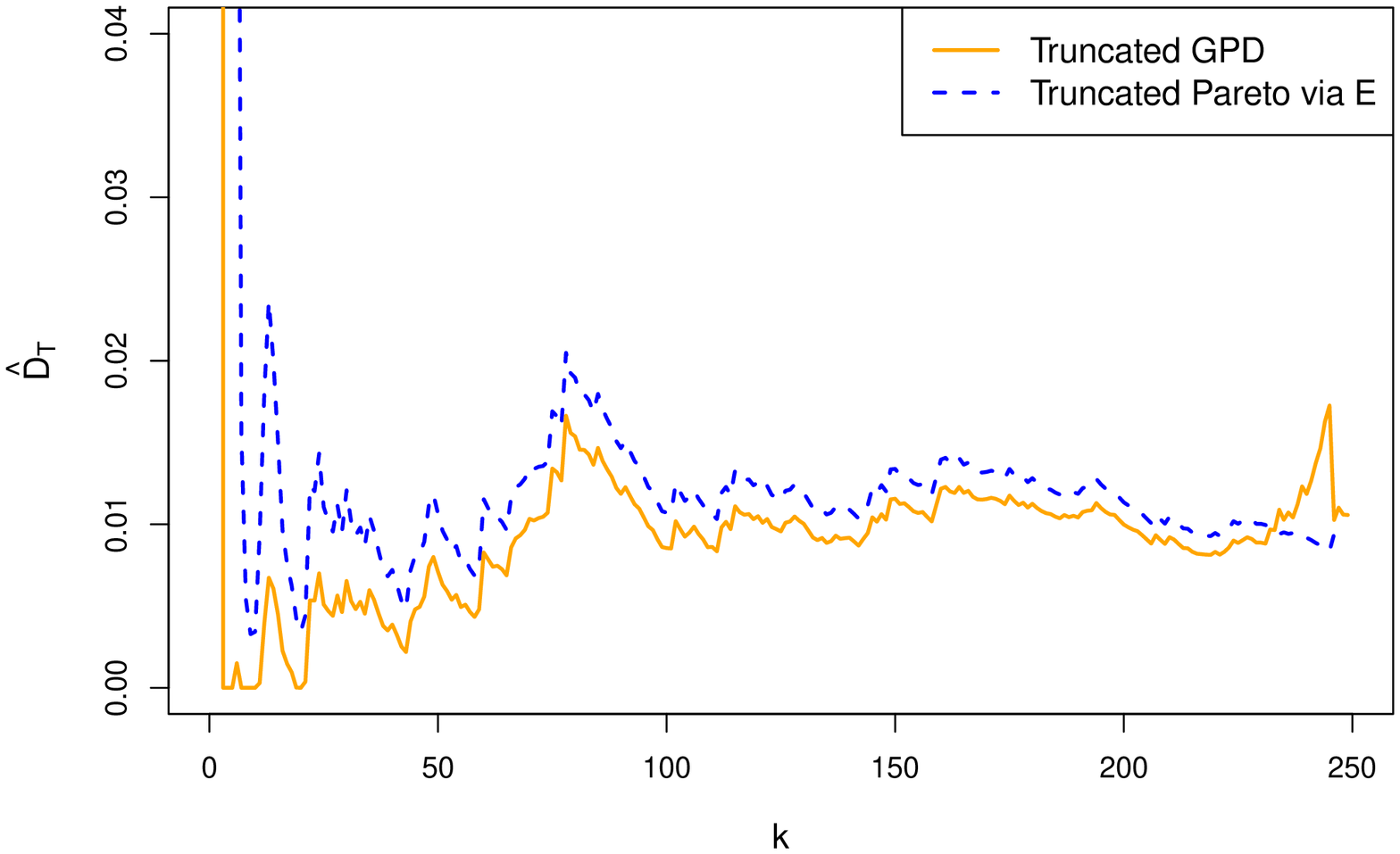}%
		\caption{}%
		\label{fig:Groningen_DT}%
	\end{subfigure}%
	\newline
	\begin{subfigure}{0.495\textwidth}
		\centering
		\includegraphics[height=\textwidth, angle=270]{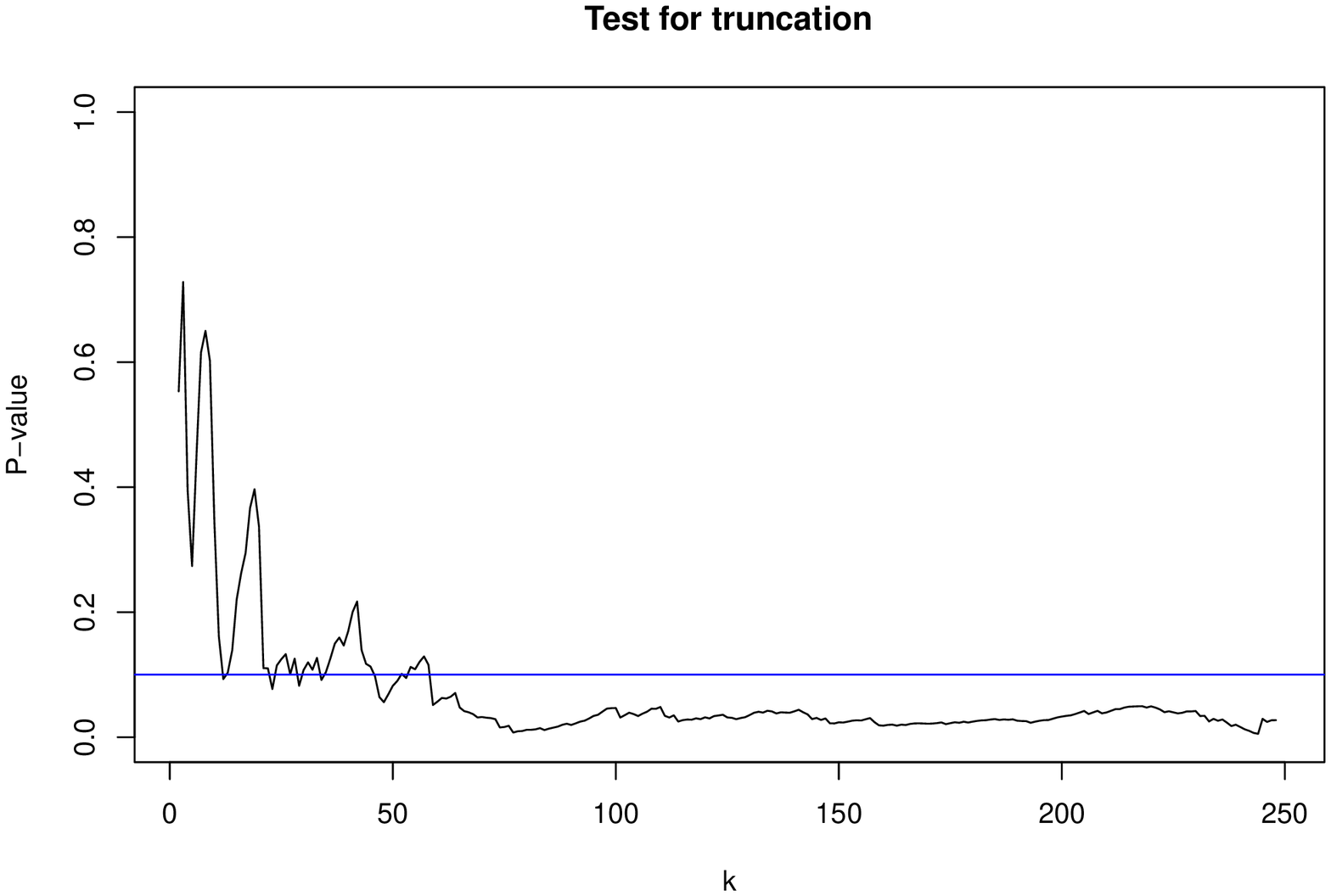}%
		\caption{}%
		\label{fig:Groningen_trTest_mle}%
	\end{subfigure}
	\begin{subfigure}{0.495\textwidth}
		\centering
		\includegraphics[height=\textwidth, angle=270]{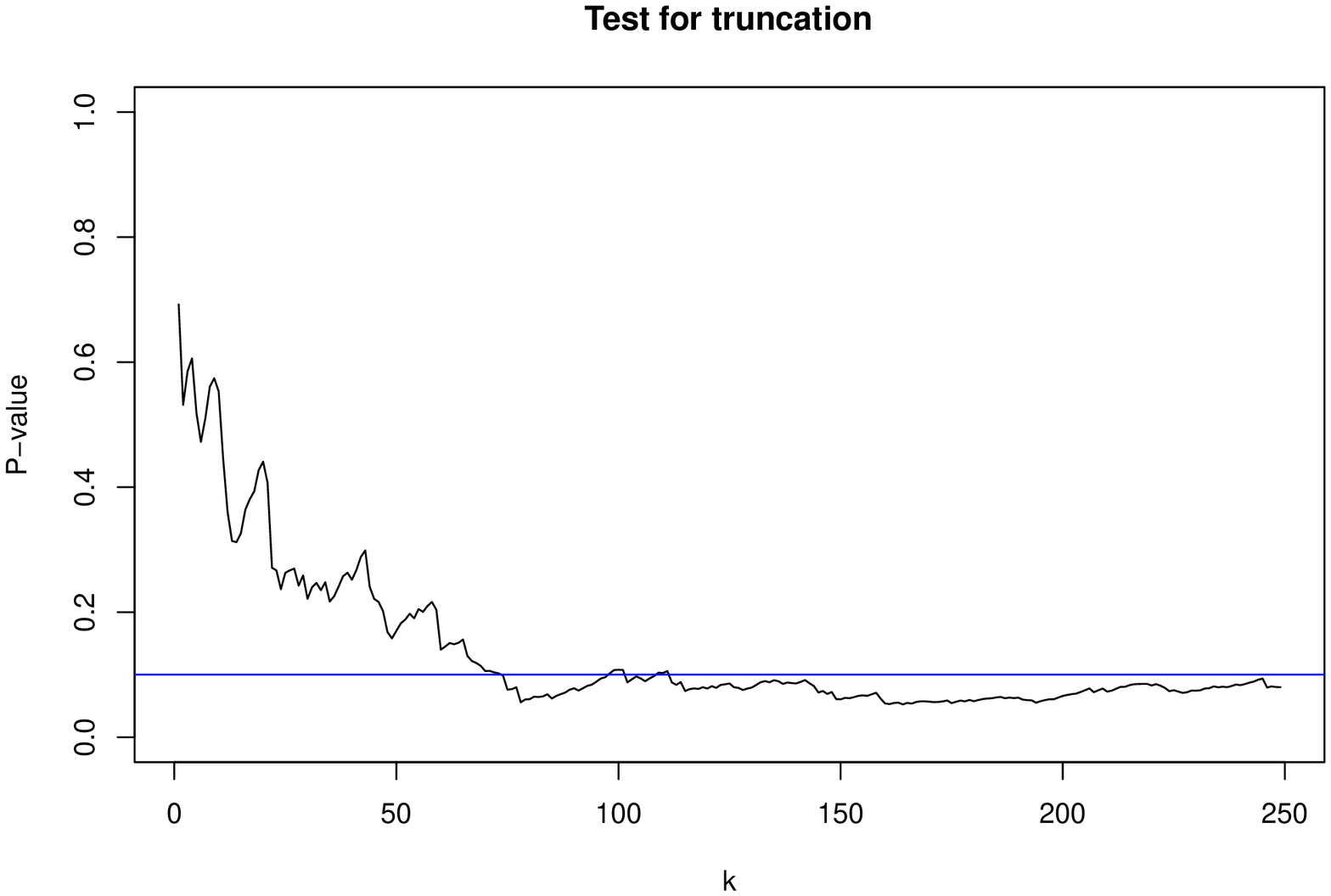}%
		\caption{}%
		\label{fig:Groningen_trTest}%
	\end{subfigure}
	\caption{GGF anthropogenic seismicity: (a) estimates of $\xi$ (full line) and $\xi_{Y_E}$ (dashed line), (b) estimates of the truncation odds $D_T$, (c) P-values of a test for truncation based on the truncated GPD and (d) P-values of a test for truncation based on the truncated Pareto.}%
\end{figure}

The magnitudes in the database are rounded to one decimal digit, and hence there are several ties in the dataset. Therefore, we smoothed the data by adding independent uniform random numbers within the range [-0.05, 0.05] to all magnitudes that occur more than once. This ensures that all observations are unique.  We then retain the 250 magnitudes larger than or equal to $t_M=1.5$. The choice of 1.5 as threshold in the Groningen case is standard in the geological literature, see e.g.\@ \citet{KNMI}. The exponential QQ-plot in Figure~\ref{fig:Groningen_expQQ} indicates that an exponential distribution is indeed suitable for the magnitudes, but the bending off at the largest observations suggests a possible upper truncated tail. The same behaviour is seen in the mean excess plot \citep[see e.g.\@ Chapter~1 in][]{SoE} in Figure~\ref{fig:Groningen_MeanExcess}: the first horizontal part suggests that the data come from an exponential-like distribution, whereas the downward trend at the end indicates an upper truncation point. Note that the Pareto QQ-plot of the energy in Figure~\ref{fig:Groningen_ParetoQQ} suggests that the energy follows a truncated Pareto distribution as discussed in Section~\ref{sec:trHill}. When applying the truncated GPD estimator to the magnitudes, a value of $\xi$ around 0 is found suggesting again an exponential-like distribution, see Figure~\ref{fig:Groningen_xi}. The parameter $\xi_{Y_E}$ is estimated by the truncated Pareto estimator to be around 1.8. The estimators for $D_T$ based on the truncated GPD and truncated Pareto estimators for $\xi$ and $\xi_{Y_E}$, respectively, suggest that the truncation odds are around 1\%, see Figure~\ref{fig:Groningen_DT}. Next, we test (directly and via the energy) if the data come indeed from an upper truncated distribution. Under the null hypotheses of both tests, the data come from an unbounded, hence not upper truncated, distribution. The P-values of a test for truncation based on the truncated GPD \citep{trunc_real} in Figure~\ref{fig:Groningen_trTest_mle} indicate, for larger values of $k$, that the magnitude data come from an upper truncated distribution. Similarly, P-values of a test for truncation based on the truncated Pareto \citep{trHill, Truncation} in Figure~\ref{fig:Groningen_trTest} indicate that, for values of $k$ above 75, the distribution of the energy is upper truncated.  Note that the significance level of the tests, 10\%, is indicated by the horizontal lines in Figure~\ref{fig:Groningen_trTest_mle} and~\ref{fig:Groningen_trTest}.
Finally, the fit provided by the truncated GPD with $k=125$, and hence $\hat{\xi}_{125}\approx0$, models the data well, see Figure~\ref{fig:Groningen_ExpQQ_fit}. All these elements suggest that the truncated Gutenberg-Richter distribution, i.e.\@ a doubly truncated exponential distribution, might indeed be a suitable model for the GGF magnitude data.

\begin{figure}[!h]
	\centering
	\begin{subfigure}{0.625\textwidth}
		\centering
		\includegraphics[height=\textwidth, angle=270]{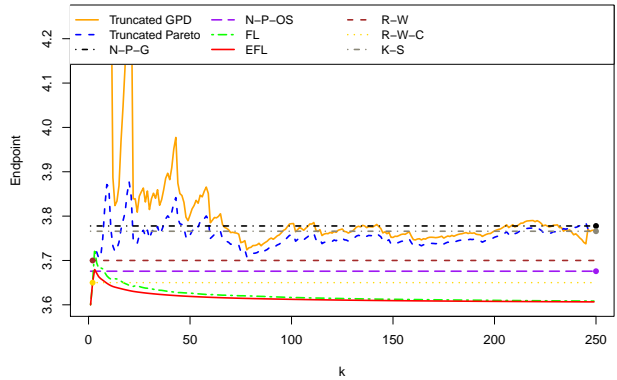}%
		\caption{}%
		\label{fig:Groningen_endpoint_all}%
	\end{subfigure}
	\hfill
	\begin{subfigure}{0.625\textwidth}
		\centering
		\includegraphics[height=\textwidth, angle=270]{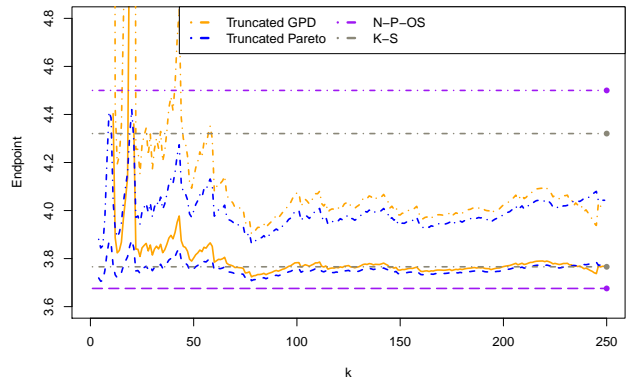}%
		\caption{}%
		\label{fig:Groningen_CI}%
	\end{subfigure}%
	\caption{GGF anthropogenic seismicity: (a) estimates of the maximum possible earthquake magnitude $T_M$ and (b) 90\% upper confidence bounds for $T_M$.}%
\end{figure}

\par Next, we compute all discussed estimates for the maximum possible earthquake magnitude (Figure~\ref{fig:Groningen_endpoint_all}). For estimators where no value of $k$ needs to be chosen, the dot indicates how many observations are used: 2 or $n$. All estimators suggest that the endpoint lies between 3.61 and 3.80 on the Richter scale. Note however, that for the estimators of the endpoint based on EVT, we need to look at larger values of $k$ where a more stable pattern emerges as the test for truncation was only significant for $k\geq 75$. For $k$ around 125, the EVT-based methods estimate the endpoint around 3.76. Note that the EVT estimates for $k=n$ are close to the estimates of the N-P-G and K-S methods which use all $n$ observations above 1.5.
All other methods lead to estimates for the endpoint that are lower than the EVT results.

\par Additionally, we look at 90\% upper confidence bounds for the endpoint as discussed above. The endpoint estimators are given by the full orange (truncated GPD), dashed blue (truncated Pareto), purple long dashed (N-P-OS) and grey dash-dotted (K-S) lines in Figure~\ref{fig:Groningen_CI}. The corresponding 90\% upper bounds are added as dash-dotted lines in the same colour. The upper bounds using the truncated GPD \eqref{eq:CI_trMLE} and truncated Pareto \eqref{eq:CI_trHill} take values of 4.04 and 3.98, respectively, for $k=125$. The 90\% upper bound \eqref{eq:CI_npOS} takes a value of 4.50, and the parametric 90\% upper bound \eqref{eq:CI_GR} is equal to 4.32 (grey point). Note that the latter two confidence bounds are based on $n$ magnitudes and should hence be compared with the EVT-based upper bounds for $k=n$ (4.03 and 4.04).

We summarised the obtained estimates and 90\% confidence bounds for the maximum possible earthquake magnitude in Table~\ref{tab:results}. Note that for the estimators where $k$ needs to be chosen, we indicate the chosen value of $k$ in the last column. Fixed values of $k$, e.g.\@ $2$ for the R-W estimator, are indicated in the last column in italics.

\begin{table}[!h]
	\centering
	\resizebox{\textwidth}{!}{  
		\begin{tabular}{l|r|r|r}
			\hline
			Estimator  & Estimated $T_M$ & 90\% upper confidence bound & $k$ \\
			\hline
			Truncated GPD & 3.77 & 4.04 & 125\\
			Truncated Pareto & 3.75 & 3.98 &  125\\
			Non-parametric Gaussian (N-P-G) & 3.78 & / & $n=$\textit{250}\\
			Non-parametric order statistics (N-P-OS) & 3.68 & 4.50 & $n=$\textit{250}\\
			Few largest observations (FL) & 3.61 & / & 250\\
			Extended few largest observations (EFL) & 3.61  & / & 250\\
			Robson -- Whitlock (R-W) & 3.70 & / & \textit{2}\\
			Robson -- Whitlock -- Cooke (R-W-C) & 3.65 & / & \textit{2} \\
			Kijko -- Sellevoll (K-S) & 3.77 & 4.32 &  $n=$\textit{250}\\
			\hline
	\end{tabular}}%
	\caption{Summary of estimates and 90\% confidence bounds for the maximum possible earthquake magnitude in the GGF.}\label{tab:results}
\end{table}

\section{Simulations}\label{sec:Groningen_sim}

The performance of the nine applied estimators of the upper limit of the magnitude distribution was tested using simulations. We generated 5000 magnitude samples of size 250 from the truncated Gutenberg-Richter distribution with level of completeness $t_M = 1.5$, rate parameter $\beta = 2.1203$ and three different endpoints: $T_M = 3.75$, 4.0 and 4.5. The parameter $\beta$ was estimated from the GGF data by application of \eqref{eq:beta_Taylor} \citep{Gibowicz_Kijko}. Note that these endpoints correspond to the 99.2\%, 99.5\% and 99.8\% quantiles of the shifted exponential distribution with $\beta = 2.1203$ and level of completeness $t_M = 1.5$. For each of these simulations, we plot the relative mean, the relative mean squared error (MSE) and the coverage percentage of the upper confidence bounds over the 5000 simulations. These plots can be found in Appendix~\ref{sec:app}.

The simulations show that the truncated GPD and truncated Pareto estimators have the lowest bias, over all three considered truncation points. However, their MSE is among the highest which indicates that these estimators have the largest variances. As expected, the bias and MSE of all nine analysed estimators increases when the endpoint gets larger. For simulations with endpoint 3.75 and 4.0 (which seem to be realistic scenarios), on average, the EVT estimators slightly overestimate the true endpoint. When $T_M = 4.5$, all estimates of $T_M$, except K-S, are on average too low. 

The coverage percentages of the upper confidence bounds are defined as the percentage of times that the obtained upper bounds are larger than the true endpoint. In theory, these percentages should be equal to 90\%. When the endpoint gets larger, the observed coverage percentages decrease. The coverage percentage for the upper bound \eqref{eq:CI_npOS} of \citet{Cooke79} is closer to 90\% than the ones for the upper bounds of the EVT-based estimators. The performance of the first two EVT-based estimators is rather similar with a slight advantage for the truncated Pareto. Since second-order bias terms were not taken into account for the upper bounds \eqref{eq:CI_trMLE} and \eqref{eq:CI_trHill}, developing bias reduced methods can improve these upper bounds. 
The parametric upper confidence bound \eqref{eq:CI_GR}, which uses all $n=250$ observations, performs similarly to the one using the truncated Pareto for $k$ large when the endpoint is 3.75. For higher endpoints, this upper confidence bound performs much worse than the other ones.

It is important to note that the parametric K-S estimator is designed specifically for the truncated Gutenberg-Richter distribution, which we consider in these simulations, whereas the EVT-based estimators are also suitable for other upper truncated distributions. The good performance of the EVT-based estimators on different upper truncated distributions, e.g.\@ a truncated lognormal distribution, is shown through simulations in \citet{Truncation, trunc_real}.

\section{Conclusions}

In our work, we investigated the performance of nine different estimators of the endpoint of the distribution, and applied it to the estimation of the maximum possible seismic event magnitude generated by gas production in the Groningen gas field. The analysis includes a comparison of EVT-based estimators, non-parametric estimators and a parametric estimator.  Since the available database contains only a few large magnitude events, all estimates provide the assessment of the upper limit of magnitude with significant uncertainty. The quantification of the uncertainty is a problem on its own, which requires careful consideration and effort, not less than the assessment of the upper limit of magnitude itself. 

Based on the application of the nine different techniques, the maximum possible anthropogenic origin seismic event magnitude in the Groningen gas field is estimated to be in the range 3.61 to 3.80. The 90\% upper confidence bounds vary from 3.85 to 4.50. In addition, the extreme value analysis in Section~\ref{sec:Groningen} suggests that the widely used truncated Gutenberg-Richter distribution might indeed be appropriate to model the distribution of seismic event magnitudes in the Groningen gas field. However, the EVT-based and non-parametric estimators do not require knowledge of the magnitude distribution, which gives them more flexibility compared to their parametric counterparts. 

Based on simulations from the truncated GR distribution, it is clear that the EVT-based methods perform well when estimating the endpoint. It is important to note that these methods usually provide an assessment with a positive bias, which means that, on average, the true endpoint is overestimated, whereas the other estimators (except K-S and N-P-G), on average, are too low. The upper confidence bounds based on these two estimators are sharper than the other ones. However, the simulations point out that they are too sharp indicating the need for bias reduction. 

In general, the presence of bias is not an obstacle leading to disqualification of any of the applied endpoint assessment procedures. It would be very useful to study the bias in detail. If we knew the bias, it could be used to correct the endpoint estimator \citep{LasockiUrban}, and potentially lead to improvement of any of the discussed procedures. Moreover, if additional, independent high-quality information is available, the Bayesian formalism provides a powerful tool, capable of both improving the endpoint estimates and providing a more reliable assessment of its confidence bounds. 

Overall, we can conclude that the EVT-based estimators of \citet{Truncation,trunc_real} are a valuable addition to the already existing methods for estimation of the area characteristic, maximum possible seismic event magnitude.


\bibliographystyle{spbasic}
\bibliography{Groningen}

\appendix

\section{Simulation results}\label{sec:app}

\begin{figure}[!h]
	\centering
	\includegraphics[height=0.625\textwidth, angle=270]{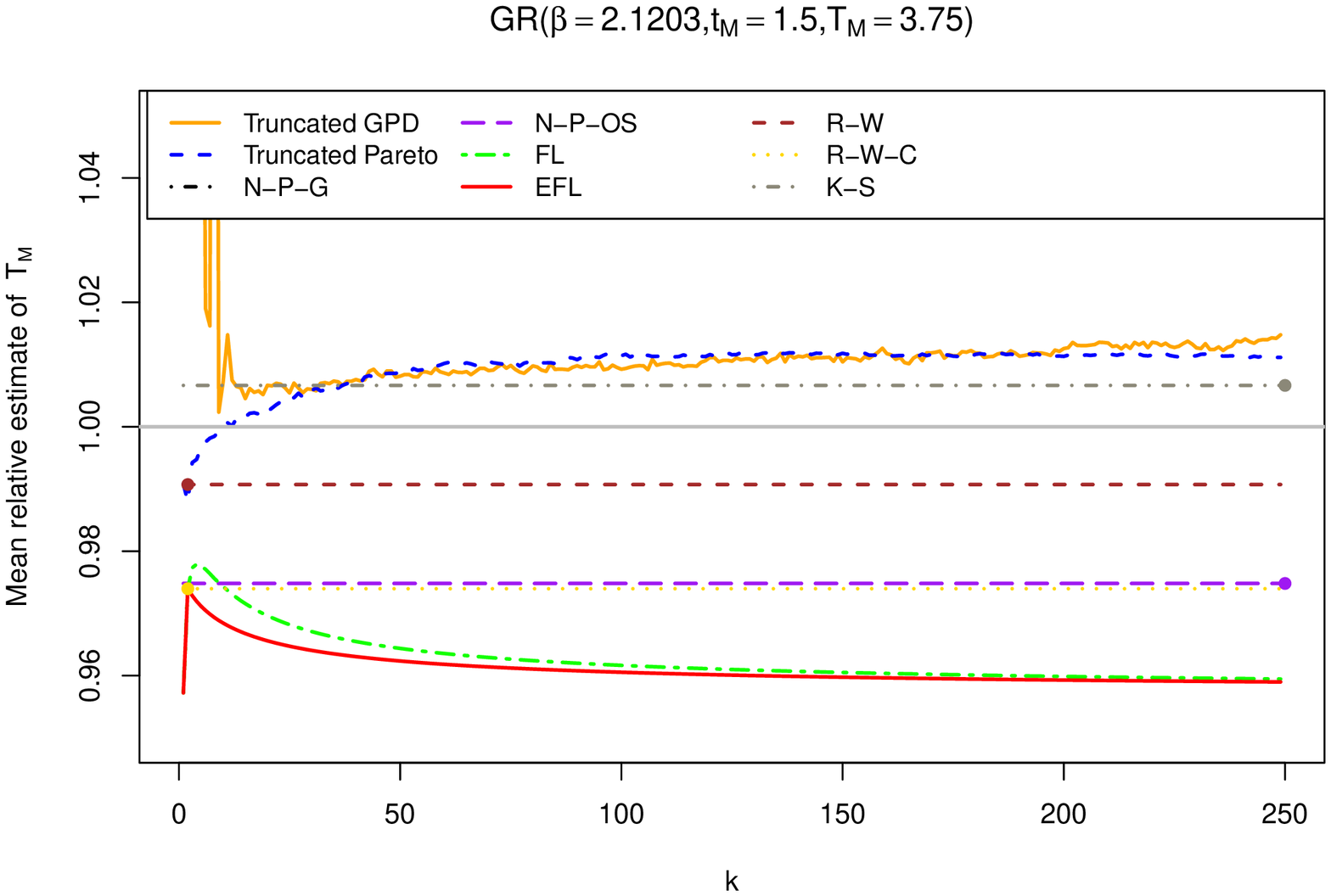}%
	
	\includegraphics[height=0.625\textwidth, angle=270]{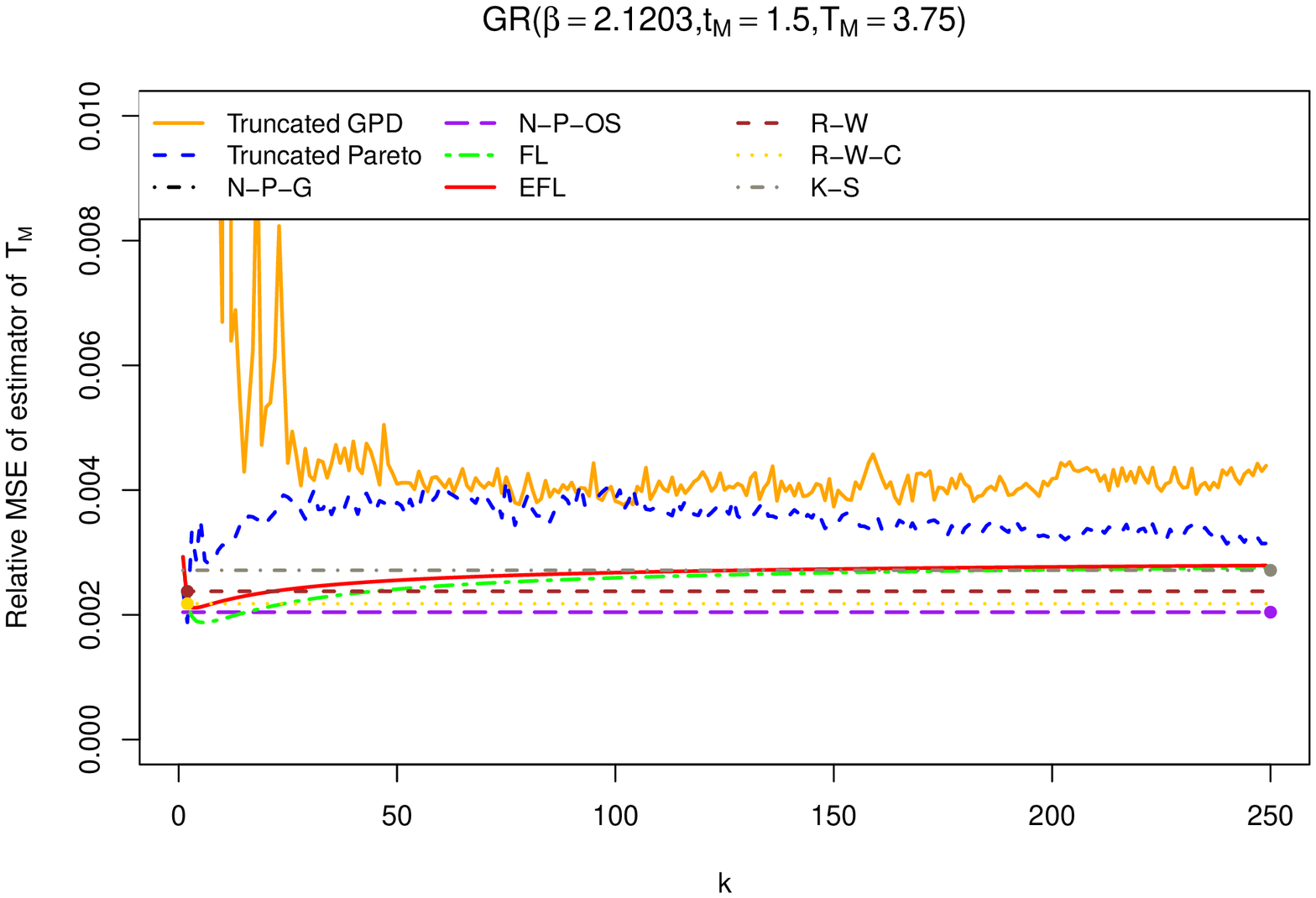}%
	
	\includegraphics[height=0.625\textwidth, angle=270]{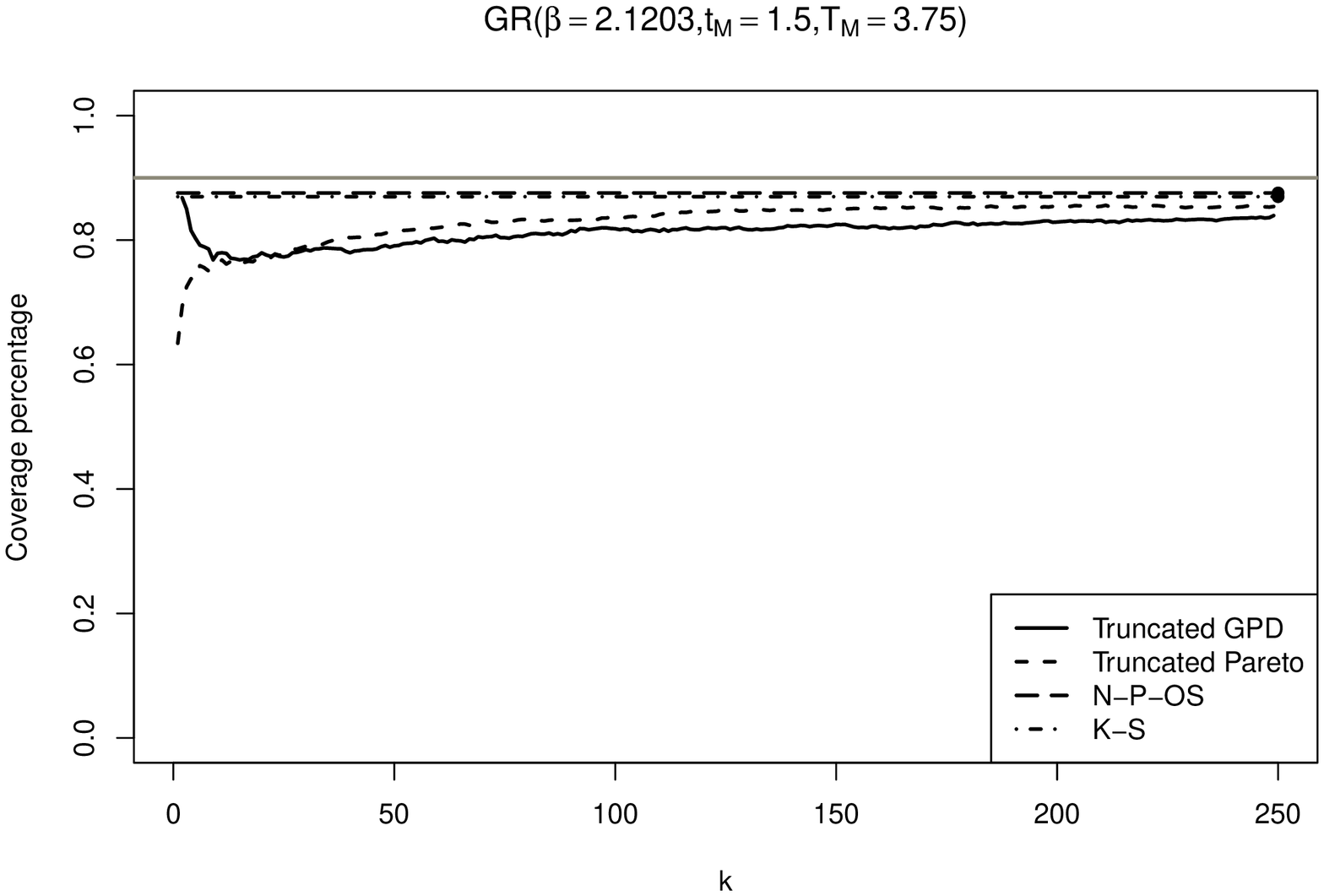}%
	\caption{$GR(\beta=2.1203, t_M=1.5, T_M=3.75)$: relative means of endpoint estimates (top), relative MSE of endpoint estimates (middle) and coverage percentage of 90\% upper confidence bounds for the endpoint (bottom).}%
\end{figure}

\begin{figure}[!ht]
	\centering
	\includegraphics[height=0.625\textwidth, angle=270]{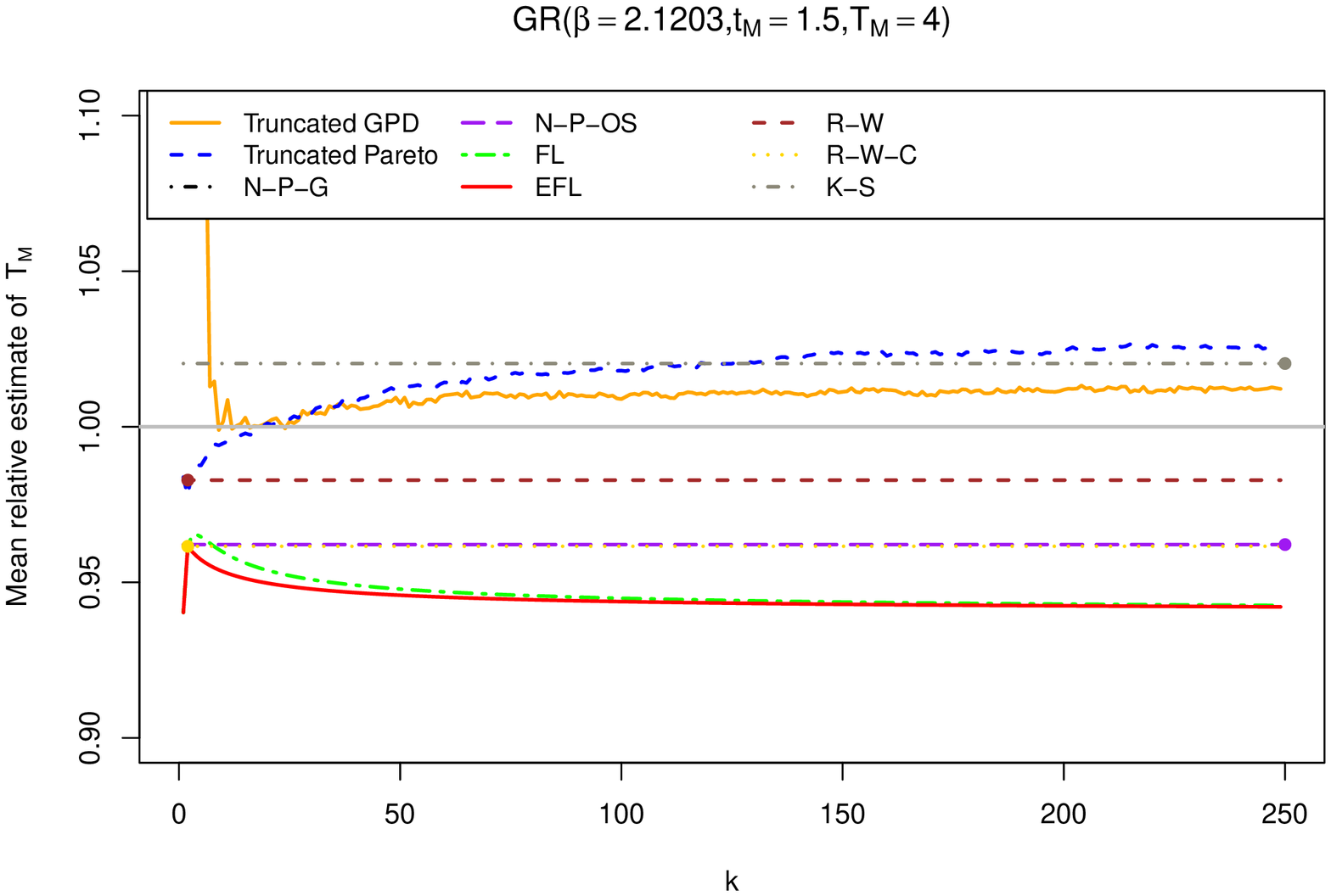}%
	
	\includegraphics[height=0.625\textwidth, angle=270]{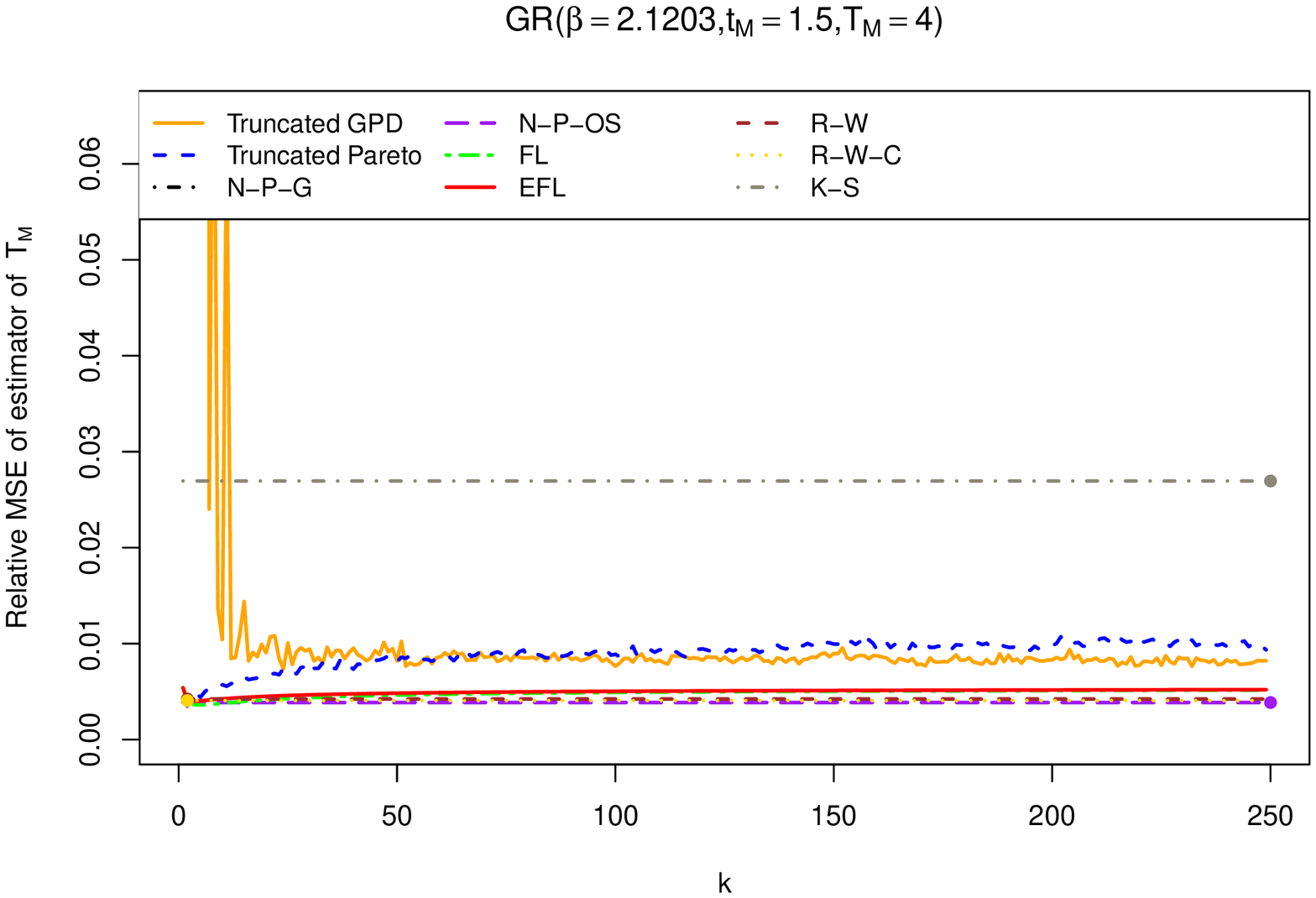}%
	
	\includegraphics[height=0.625\textwidth, angle=270]{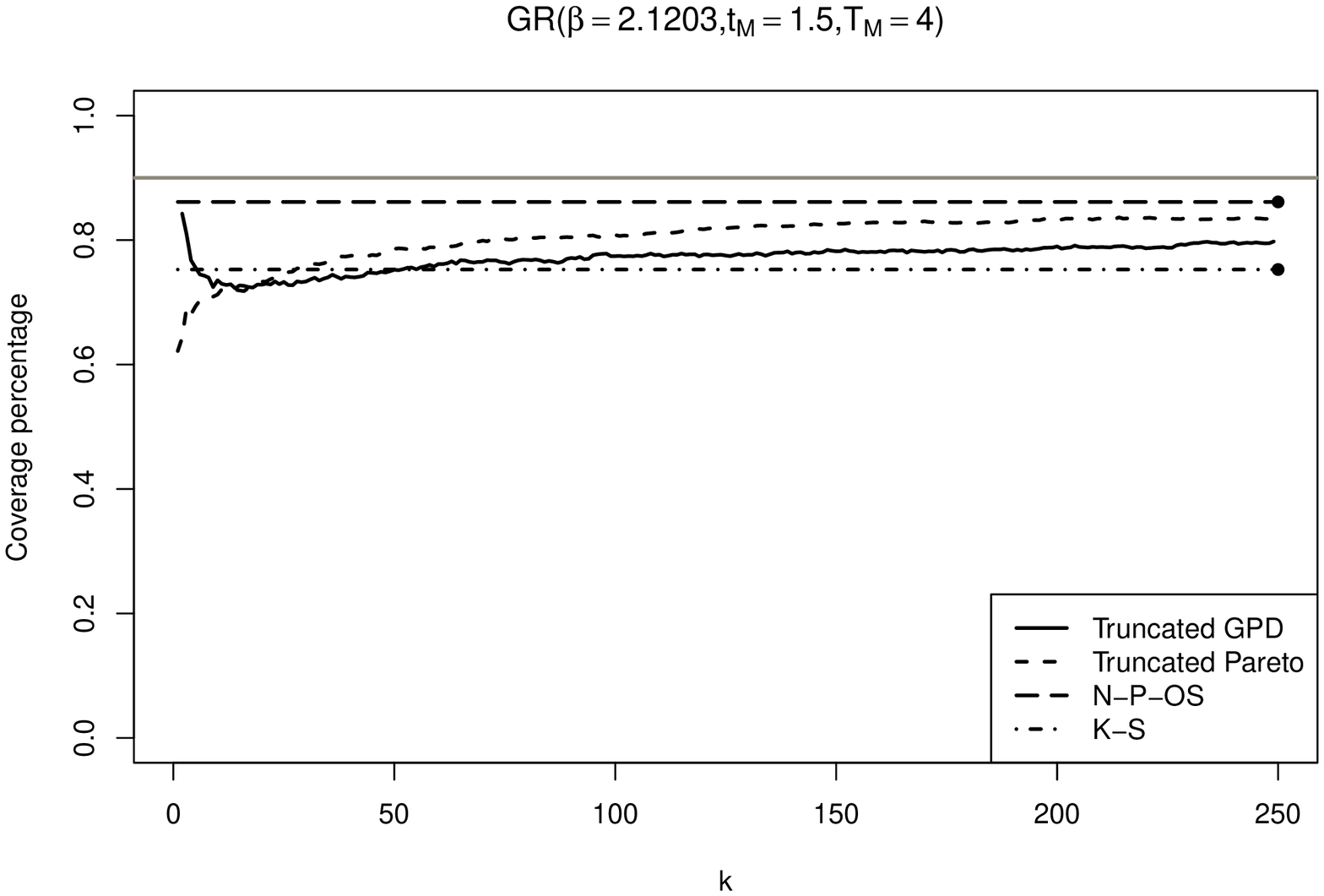}%
	\caption{$GR(\beta=2.1203, t_M=1.5, T_M=4)$: relative means of endpoint estimates (top), relative MSE of endpoint estimates (middle) and coverage percentage of 90\% upper confidence bounds for the endpoint (bottom).}%
\end{figure}

\begin{figure}[!ht]
	\centering
	\includegraphics[height=0.625\textwidth, angle=270]{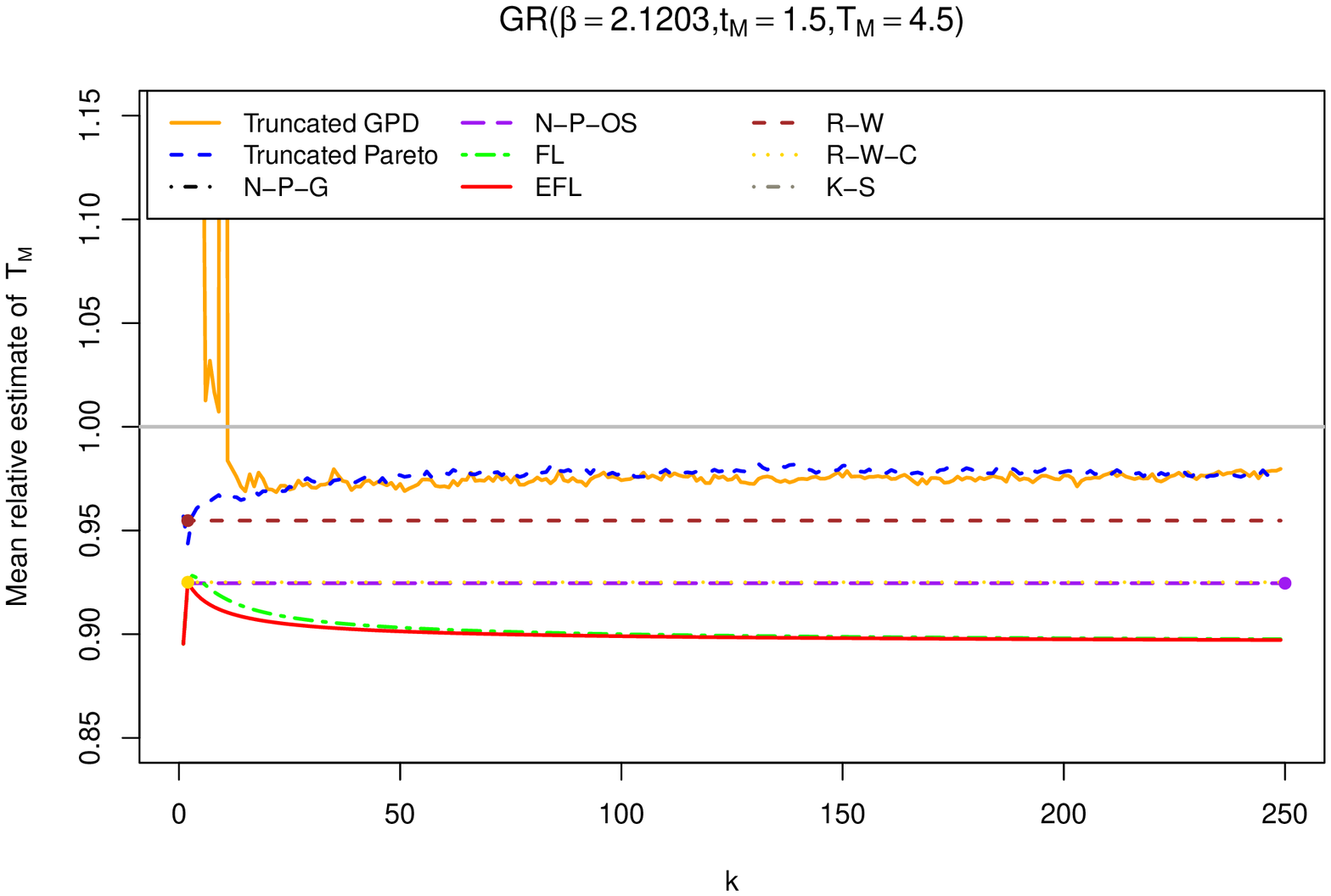}%
	
	\includegraphics[height=0.625\textwidth, angle=270]{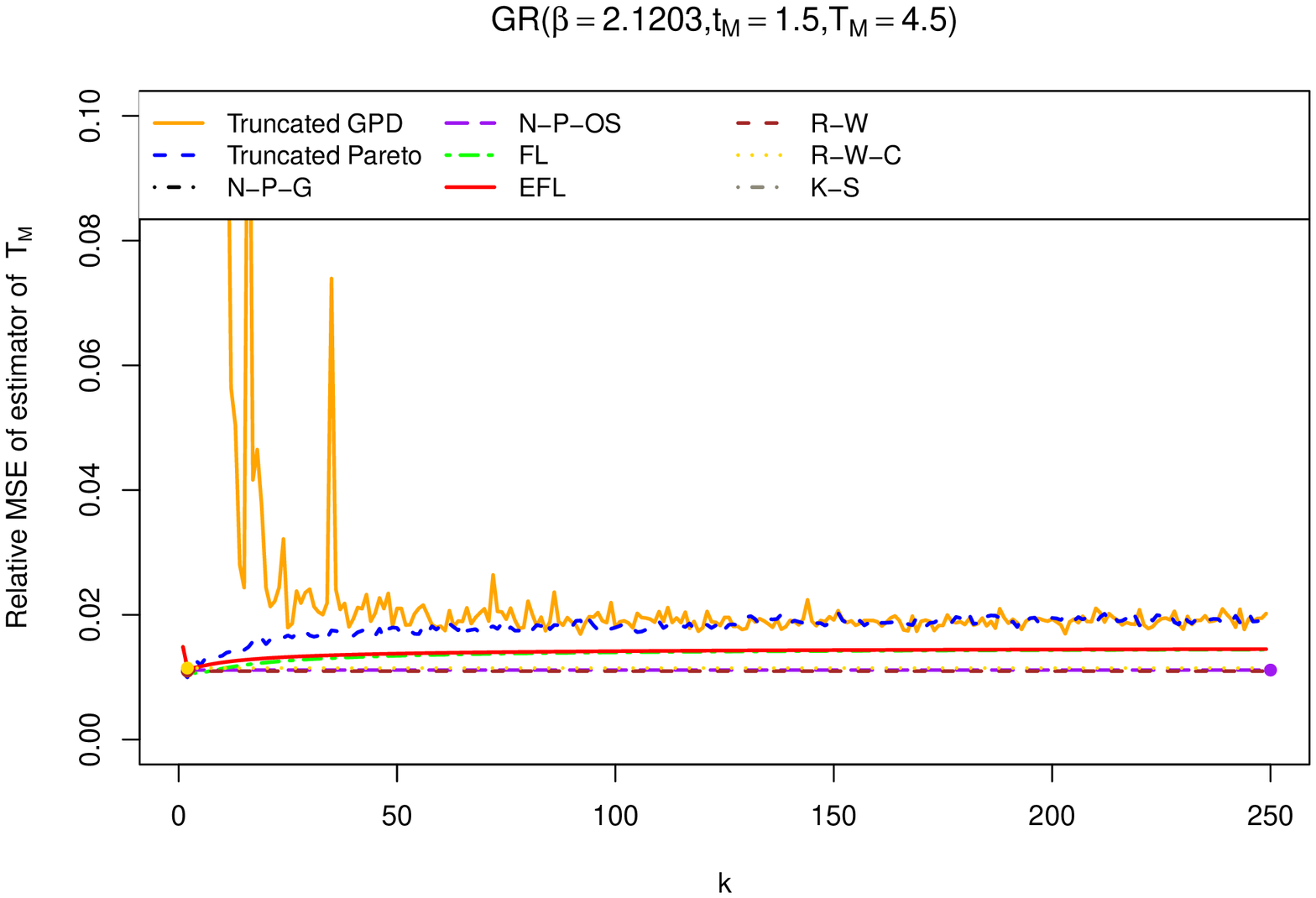}%
	
	\includegraphics[height=0.625\textwidth, angle=270]{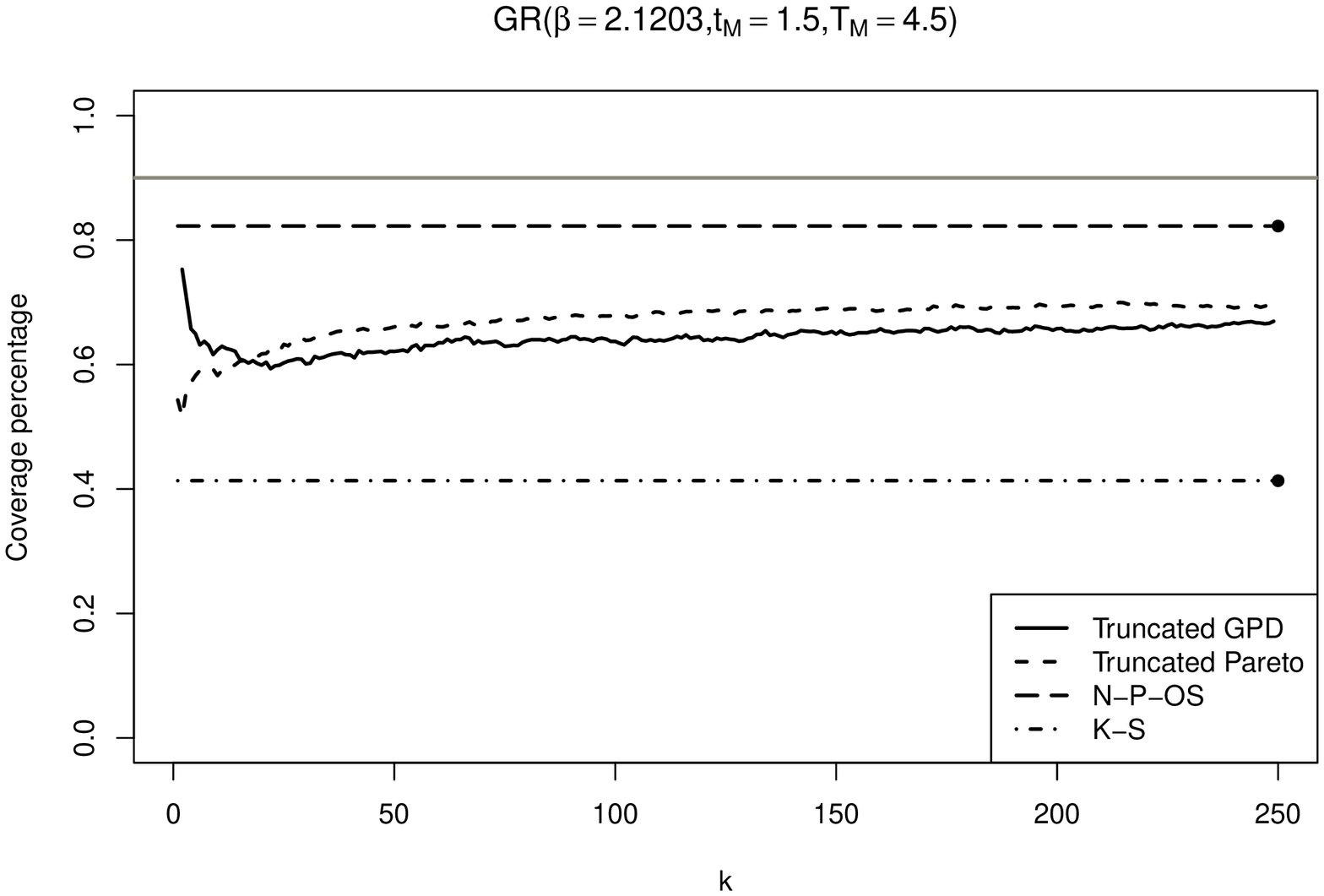}%
	\caption{$GR(\beta=2.1203, t_M=1.5, T_M=4.5)$: relative means of endpoint estimates (top), relative MSE of endpoint estimates (middle) and coverage percentage of 90\% upper confidence bounds for the endpoint (bottom).}%
\end{figure}
\end{document}